\documentclass[12pt]{article}
\usepackage{jheppub_my} 

\usepackage{amsmath,amssymb,amsfonts}
\usepackage{graphicx}
\graphicspath{{./Figure_orig/}}

\newcommand {\beq}{\begin{eqnarray}}
\newcommand {\eeq}{\end{eqnarray}}

\newcommand{\bC}{\ensuremath{\mathbb{C}}}

\newcommand{\bH}{\ensuremath{\mathbb{H}}}

\newcommand{\bR}{\ensuremath{\mathbb{R}}}

\newcommand{\bZ}{\ensuremath{\mathbb{Z}}}

\newcommand{\scA}{\ensuremath{\mathcal{A}}}

\newcommand{\scD}{\ensuremath{\mathcal{D}}}

\newcommand{\scF}{\ensuremath{\mathcal{F}}}
\newcommand{\scG}{\ensuremath{\mathcal{G}}}
\newcommand{\scH}{\ensuremath{\mathcal{H}}}

\newcommand{\scL}{\ensuremath{\mathcal{L}}}
\newcommand{\scM}{\ensuremath{\mathcal{M}}}
\newcommand{\scN}{\ensuremath{\mathcal{N}}}

\newcommand{\scT}{\ensuremath{\mathcal{T}}}

\newcommand{\scW}{\ensuremath{\mathcal{W}}}

\newcommand{\Teichmuller}{Teichm\"{u}ller }
\newcommand{\Kahler}{K\"ahler }

\newcommand{\sfX}{\ensuremath{\mathsf{X}}}
\newcommand{\sfY}{\ensuremath{\mathsf{Y}}}
\newcommand{\sfZ}{\ensuremath{\mathsf{Z}}}
\newcommand{\sfT}{\ensuremath{\mathsf{T}}}
\newcommand{\sfx}{\ensuremath{\mathsf{x}}}
\newcommand{\sfy}{\ensuremath{\mathsf{y}}}
\newcommand{\sfz}{\ensuremath{\mathsf{z}}}

\newcommand{\sfL}{\ensuremath{\mathsf{L}}}
\newcommand{\sfR}{\ensuremath{\mathsf{R}}}
\newcommand{\sfr}{\ensuremath{\mathsf{r}}}
\newcommand{\sfs}{\ensuremath{\mathsf{s}}}

\title{\boldmath $SL(2,\bR)$ Chern-Simons, Liouville, \\
and Gauge Theory on Duality Walls}

\author[\spadesuit]{Yuji Terashima}
\author[\clubsuit]{and Masahito Yamazaki}

\affiliation[\spadesuit]{Department of Mathematics, Tokyo Institute of Technology, Tokyo 152-8551, Japan}
\affiliation[\clubsuit]{Princeton Center for Theoretical Science, Princeton University, 
NJ 08544, USA}
\dedicated{Dedicated to the victims of the 2011 Japan earthquake and tsunami}

\abstract{
We propose an equivalence of the partition functions of two different
3d gauge theories. On one side of the correspondence we consider the
partition function of 3d $SL(2,\bR)$ Chern-Simons theory on a
3-manifold, obtained as a punctured Riemann surface times an interval.
On the other side we have a partition function of a 3d $\scN=2$
superconformal field theory
on $S^3$, which is realized as a duality domain wall in a 4d gauge theory on $S^4$.
We sketch the proof of this conjecture using connections with quantum
Liouville theory and quantum \Teichmuller theory, and study in detail
the example of the once-punctured torus.
Motivated by these results we advocate a direct Chern-Simons
interpretation of the ingredients of (a generalization of) the
Alday-Gaiotto-Tachikawa relation. We also comment on 
M5-brane realizations as well as on possible generalizations of our proposals.
}

\begin{document}
\maketitle

\section{Introduction} \label{sec.intro}

Over the past decades the notion of duality has played a key role in
the study of gauge theories and string theories. 
The word duality traditionally refers to an equivalence of 
two different physical {\it theories}, but
there are important examples of ``duality'' which do not fall into this
category: a pair of two theories, although different as physical
theories, share certain common quantities,
which have different physical meanings and can be computed by completely
different
methods in the two theories.
In other words, this is a duality as an equivalence of two
{\it quantities} (e.g. the partition functions) or {\it subsectors}
(e.g. ground state Hilbert spaces) of the two theories, as opposed to an
equivalence of
the full theories themselves.
This extended notion of duality, although limited
in its power and scope, is much more general, and provides a
useful bridge between theories which are unrelated otherwise.
A good example for this is the recently discovered
Alday-Gaiotto-Tachikawa (AGT) relation, which claims an equivalence of the instanton
partition functions of 4d $\scN=2$ $SU(2)$ superconformal 
field theories on $S^4$ and conformal blocks
 of 2d Liouville theory on Riemann surfaces \cite{Alday:2009aq}. 

In this paper, we initiate a program to connect two different 3d gauge
theories --- one is a (bosonic) 3d $SL(2,\bR)$ Chern-Simons theory and
another is a 3d $\scN=2$ superconformal field theory.
This provides yet another example of the generalized duality in the
sense mentioned above. 
In particular, we propose an equivalence of the partition function of
the two theories. This is schematically written as
\beq
Z_{\textrm{3d $SL(2,\bR)$ CS}}=Z_{\textrm{3d $\scN=2$ theory}}.
\label{Z=Zintro}
\eeq
More precise formulation of this statement, as well as explicit
examples, will be given below.

We will arrive at the relation \eqref{Z=Zintro} by a chain of
connections with quantum \Teichmuller theory and quantum Liouville
theory, each step interesting in its own right.
For example, in one step we will encounter 
a new state sum model for $SL(2,\bR)$
Chern-Simons theory defined from quantum \Teichmuller theory.
Assuming several conjectures in the literature, this almost gives a proof
of the relation \eqref{Z=Zintro}.
This will also clarify the connection between our proposal 
and the AGT conjecture and its generalization \cite{Drukker:2010jp}.
From this viewpoint the relation \eqref{Z=Zintro} should be thought of 
as a (3+3) version of AGT relation, which divides 6 into (4+2) (see the
cautionary remarks in section \ref{sec.idea}, however).
In fact, this understanding leads to an interesting question of reformulating everything we
know about the AGT correspondence (and its generalization) in the language of $SL(2,\bR)$
Chern-Simons theory. We will provide partial answer to this problem in this
paper. In particular, the underlying reason for our correspondence is that there is a natural
identification of the Hilbert spaces of all the four theories, which
identification one might
be tempted to call ``Gauge/Liouville/Teichm\"uller/Chern-Simons
correspondence''\footnote{As we
will discuss later, for the last equality concerning \Teichmuller theory and
Chern-Simons theory we need to specify an appropriate boundary condition for
the Chern-Simons theory.}
\beq
\scH_\textrm{4d gauge}(S^3)=\scH_\textrm{Teichm\"uller}(\Sigma)=\scH_{\rm Liouville}(\Sigma)=\scH_\textrm{Chern-Simons}(\Sigma).
\eeq

The rest of this paper is organized as follows. In section \ref{sec.summary} we
give more precise formulation of our conjectures. In section
\ref{sec.idea}
we outline the logic
used in deriving the conjecture.
Section \ref{sec.example} discuss the example of once-punctured torus in detail.
In section \ref{sec.higher} we briefly comment on generalizations to higher
rank gauge groups. In section \ref{sec.discussion} we discuss several
open problems which we hope will be answered in the future. 
In appendices we collect results useful for the understanding of this
paper. In particular, readers are encouraged to consult appendix
\ref{sec.Teichmuller} for the construction of the
Hilbert space in quantum \Teichmuller theory.

\section{Summary of Our Proposal}  \label{sec.summary}

In order to give a more precise formulation of \eqref{Z=Zintro}, let us explain the necessary ingredients in detail. 
On the left hand side of the correspondence we consider a Chern-Simons
theory with a non-compact gauge group $SL(2,\bR)$ \footnote{In this paper
we do not distinguish a Lie algebra from a Lie group, and use the same
symbol for both of them.}. This theory has a Lagrangian
\beq
S=\int_{M} \frac{k}{4\pi} \textrm{Tr}\left( \scA \wedge
d\mathcal{A}+\frac{2}{3} \scA\wedge \scA\wedge \scA
\right)
,
\label{SL2Raction}
\eeq
where $\scA$ is a $SL(2,\bR)$-valued connection, 
$M$ is a 3-manifold, and $k$ is the level of the
Chern-Simons theory. 
The partition function is defined by
\beq
Z(M)=\int \!\scD\! \scA\,\, e^{i S}.
\eeq
When the 3-manifold $M$ has boundaries, we need to supplement this
definition with the choice of boundary conditions.

For the purpose of this paper our 3-manifold $M$ will be defined from the following three ingredients. First, we choose a Riemann surface $\Sigma_{g,h}$, where $g$ ($h$) denotes the genus (the number of holes). $\Sigma_{g,h}$ is hereafter often going to be written $\Sigma$ for notational simplicity. We assume that $\Sigma$ is hyperbolic, i.e., satisfies
\beq
2g-2+h>0.
\label{hyperboliccond}
\eeq
Another technical subtlety here is that the holes in general are assumed
to be of finite size, i.e. each puncture $p_i$ is associated with a
number $m_i$, representing the size of the puncture\footnote{In the literature, a distinction is
sometimes made between a ``hole'' and a ``puncture'', the former having a finite
size and the latter zero size. In this paper we use both words
interchangeably, but the readers should keep in mind that the
holes/punctures are in general of finite size.}.

Second, we choose
two boundary conditions on the Riemann surface $\Sigma$.
As we will see later, for boundary conditions we first need to choose a pants decomposition of $\Sigma$
and then a set of $3g-3+h$ integers,
which we collectively write $l$ (and $l'$ for another).
We choose the same pants decomposition for the two boundary conditions $l$ and $l'$.
$l$ and $l'$ will be the length coordinates, half of the length-twist
coordinates of the \Teichmuller space of $\Sigma$ (see appendix \ref{subsec.length}).
Recall that the \Teichmuller space $\scT_{g,h}$ of a Riemann surface $\Sigma_{g,h}$
is the space of complex structure deformations of $\Sigma_{g,h}$, divided by the identity component of the diffeomorphisms of $\Sigma_{g,h}$:
\beq
\scT_{g,h}=\frac{\{ \textrm{complex structure on  } \Sigma_{g,h}\}
}{\mathrm{Diff}_0(\Sigma_{g,h})}.
\label{Teichmullerdef}
\eeq
This is a \Kahler manifold of complex dimension
\beq
\mathrm{dim}_{\bC}\,\scT_{g,h}=
3g-3+h.
\label{dimTgh}
\eeq

Third, we fix an
element $\varphi$ of the mapping class group $\Gamma_{g,h}$ of
$\Sigma$, which is defined by
\beq
\Gamma_{g,h}=\frac{\mathrm{Diff}(\Sigma_{g,h})}{\mathrm{Diff_0}(\Sigma_{g,h})}.
\label{MCG}
\eeq
This acts on $\scT_{g,h}$, and the quotient $\scM_{g,h}=\scT_{g,h}/\Gamma_{g,h}$ is the moduli space of the Riemann surface.

Our 3-manifold $M_{\Sigma, (l,l'), \varphi}$ is then defined to be
$\Sigma\times I$, where $I=[0,1]$ is an interval and the boundary
conditions at $\partial M=\Sigma\times \{0\} \cup \Sigma\times \{1\}$
are determined by $l$ and $\varphi(l')$ \footnote{Note that $l$ and
$l'$ have the same pants decomposition, whereas $\varphi(l')$ in general
does not.}.
The Chern-Simons partition function we have in \eqref{Z=Zintro} is
defined on this manifold. When we make the dependence explicit, we have 
\beq
Z_{CS}\left[M_{\Sigma, (l,l'), \varphi}\right](m,k),
\eeq
where $m$ collectively refers to a set of parameters $\{m_i\}$.

\bigskip

Let us next describe the other side of the correspondence. On this side
we regard the Riemann surface $\Sigma$ as the so-called Gaiotto curve
\cite{Gaiotto:2009we}. In other words we consider 4d $\scN=2$
generalized quiver superconformal field theories obtained by compactifying 6d $(2,0)$ theory
(theories on multiple M5-branes) on $\Sigma$. For the most of this paper
we consider the case of two M5-branes, i.e. the gauge group of the 4d
theory is $SU(2)$. We will briefly comment on the higher rank generalization in
section \ref{sec.higher}. 

In Gaiotto's construction, the moduli space of the Riemann surface
$\Sigma$ is
interpreted as the space of marginal deformations of 4d superconformal
field theory, and the mapping class group of the Riemann surface is
identified with the S-duality group of the 4d $\scN=2$ theory.
This in particular means
that we can regard $\varphi$ as a symmetry of the 4d $\scN=2$ theory,
and the physics at value of complexified gauge coupling $\tau$
 is
equivalent to the physics at $\varphi(\tau)$. 
Here $\tau$ is defined from the gauge coupling constant $g$ and the
theta-angle $\theta$ by
\beq
\tau=\frac{4\pi i}{g^2}+\frac{\theta}{2\pi}.
\eeq
The 4d theory is in
general strongly coupled -- only at the boundary of the moduli space 
where the Riemann surface degenerates (this is specified by a choice of
pants decomposition) do we have a Lagrangian
description of the 4d theory.
Finally, the hole parameters $m_i$ in 2d 
are interpreted as the mass parameters of the 4d theory,

The 3d $\scN=2$ theory we are interested in is realized as a theory on the
1/2 BPS duality domain wall inside this 4d $\scN=2$ theory. To define
this, let us consider $\bR^{3,1}$, and divide one of the spatial
directions (say, $x^3$) into two parts, $x^3>0$ and $x^3<0$. On one side
$x^3>0$ we consider 4d theory with complexified gauge coupling $\tau$,
and on the other ($x^3<0$) we consider the same theory with different
values of the complexified gauge coupling: we take the value to be
$\varphi(\tau)$, where as above $\varphi$ is an element of the duality
group. The complexified gauge coupling $\tau$ then has a non-trivial
profile near $x^3=0$ (Janus solution), and we in particular consider a
profile preserving half of the supersymmetries.
We can make the value of $\tau$ constant in the whole $\bR^{3,1}$ by
taking an S-duality in $x^3<0$, and all the effect of $\varphi$ is 
localized around $x^3=0$ \footnote{There is a subtlety here. When we
take the S-dual of $SU(2)$ gauge theory, the gauge group becomes the
Langlands dual $SU(2)^\vee=SO(3)$, which is different from the
original gauge group $SU(2)$ by a discrete gauge group $\bZ_2$.
In this paper we only deal with Lie algebras, and neglect such global
structures of Lie groups.}.
 Since our 4d theory is
conformal we can squeeze everything into $x^3=0$ and we
have a 3d domain wall at $x^3=0$. This domain wall is often called
a ``duality wall'' since it is specified by an element $\varphi$ of the
4d duality group. It is the 3d $\scN=2$ theory (1/2 BPS in 4d $\scN=2$
theory)
on this duality domain wall 
that we study in this paper. We are going to call the 4d theory the
mother theory, and the 3d theory the daughter theory.
In the following we will denote the daughter theory by $T[SU(2); \varphi; m]$ \footnote{This is a generalization of the notation of \cite{Gaiotto:2008ak}. Their $T[SU(2)]$ will be our $T[SU(2); \varphi=S; m=0]$ for $\Sigma=\Sigma_{1,1}$.}.

In the above description of the 3d theory, 3d theory on the wall couples
to the bulk 4d theory. However, as we will see later in examples, 3d
theories themselves can be defined purely in 3d; the bulk gauge
fields couple with the 3d theory by gauging the global symmetries in
3d. The 3d theory has global symmetry $SU(2)\times SU(2)$ \footnote{As we will see, this can actually be $SU(2)\times
 SU(2)^{\vee}$ or $SU(2)^\vee\times SU(2)^{\vee}$, where
$SU(2)^\vee=SO(3)$ is the Langlands dual of $SU(2)$.}, and
correspondingly the theory has two parameters $a$ and $a'$, each
representing 
either the
mass parameter or 
the Fayet-Iliopoulos (FI) parameter. These parameters are defined by
coupling 3d theory to a background gauge fields.

It should be emphasized that our daughter theories can be defined purely
in 3d, without the coupling to the bulk 4d theory.
 The reason we refer to 4d gauge theories is twofold: first, very little
 is known about these 3d theories themselves (except for the case of
 $\Sigma_{1,1}$ which will be discussed later). Second, our definition
 of 3d theory as a theory of domain walls in 4d will be crucial when we
 discuss the connection with the AGT relation.

Let us make one cautionary remark here. This 3d theory we just described
often (but not always) contains gauge fields with Chern-Simons
terms with compact gauge group $SU(2)$. This Chern-Simons term should not
be confused with the pure bosonic $SL(2,\bR)$ Chern-Simons theory in the
other 3d. For the most of the following discussion the word Chern-Simons
 often refers to the latter theory.
 
Finally, the right hand side of
\eqref{Z=Zintro} is defined as a partition function of the 3d theory on
a $S^3$, whose metric is deformed from the
standard metric by a single parameter $b$. 
In this paper we call this 3-sphere a deformed 3-sphere, and denote it
by $S^3_b$ \footnote{In the literature this is sometimes called a
squashed $S^3$. However, readers should keep in mind that $S^3_b$
preserves only $U(1)\times U(1)$, whereas squashed 3-sphere often refers to a
3-sphere with $SU(2)\times U(1)$ isometry.}.
The deformed 3-sphere $S^3_b$ preserves $U(1)\times U(1)$ isometry of
the $SO(4)=SU(2)\times SU(2)$ isometry of the standard $S^3$, and 
can be defined by an equation
\beq
b^2 (dx_1^2+dx_2^2)+b^{-2} (dx_3^2+dx_4^2)=1.
\label{squashedmetric}
\eeq
When set to $b=1$, this reduces to the $S^3$ with the standard metric.

Now we can define the partition function of our theory on $S^3_b$
\beq
Z_{\left[  T[SU(2); \varphi; m] \right]}\left[S^3_b\right]\left( a, a'\right).
\eeq
The \eqref{Z=Zintro} is an equivalence of two expressions given so far, i.e.,
\beq
Z_{\left[  T[SU(2); \varphi; m] \right]}\left[S^3_b\right]\left( a, a'\right)
=Z_{CS}\left[M_{\Sigma, (l,l'), \varphi}\right] (m,k),
\label{main1}
\eeq
under the parameter identification
\beq
a=l, \quad a'=l',
\eeq
and
\beq
\hbar:=\frac{4\pi}{k+2}=2\pi b^2.
\label{kb}
\eeq

\bigskip

In the discussion above we have chosen a 3-manifold $M_{\Sigma,l,\varphi}$
with two boundaries. Instead, we can choose to close up the boundaries
by identifying the two boundary Riemann surfaces.
We then have the mapping torus $M_{\Sigma, \varphi}$, which is defined to be a $\Sigma$-bundle over $S^1$, with an action of $\varphi$ when we go around $S^1$:
\beq
M_{\Sigma,\varphi}=(\Sigma \times \left[0,1 \right])/\{(x,0)\sim
(\varphi(x),1)\}.
\label{mappingtorus}
\eeq
There is a corresponding operation on the gauge theory side. Our gauge
theory, $T[SU(2); \varphi; m]$, is defined by a linear quiver with
two ends, and we can identify the two ends of the quiver to make it into
a circular quiver (see Figure \ref{STfigure} (b) and (c)). The dependence of the partition function on $a$ and
$a'$ are drop out (these parameters are going to be integrated out), and \eqref{main1} becomes a statement 
\beq
Z_{CS}\left[M_{\Sigma,\varphi}\right](m,k)=
 Z_{T[SU(2); \varphi; m]}\left[S^3_b\right].
\label{main2}
\eeq

\bigskip

The parameter identification \eqref{kb} is highly non-trivial.
First, the value of $k$ is quantized\footnote{The gauge field is in general not a
globally defined 1-form, but a connection of a line bundle. 
\eqref{SL2Raction} is an expression at the local patch 
when we specify a
trivialization of the line bundle. The level is quantized in order to
make the action independent of the choice of this trivialization.}
while $b$ is a continuous
parameter. 
This suggests
that on the Chern-Simons side it is natural to perform an analytic
continue the gauge group into the
complexified gauge group $SL(2,\bC)$, whose action reads
\beq
\!\! S=\frac{t}{8\pi}\! \int \!\textrm{Tr} \left(\scA\wedge d\scA+
\frac{2}{3}\scA \wedge \scA\wedge \scA\right)
+\frac{\bar{t}}{8\pi}\! \int \!\textrm{Tr}\left( 
\bar{\scA}\wedge d\bar{\scA}+\frac{2}{3} \bar{\scA}\wedge
\bar{\scA}\wedge \bar{\scA}
\right)\! , \,
\eeq
where $\scA$ ($\bar{\scA}$) are holomorphic (antiholomorphic) connection
and we write $t=k+s, \bar{t}=k-s$. Consistency requires that $k$
is an integer as usual, but $s$ is not, and can either be real or pure
imaginary \cite{Witten:1989ip}. Such an analytic continuation is natural
also from the viewpoint of 6d (2,0) theory, since a triplet of Higgs
scalars coming from the dimensional reduction of a 6d vector multiplet 
complexifies the gauge field (cf. \cite{Dimofte:2010tz}). 

Second, there is a symmetry 
\beq
b \leftrightarrow b^{-1}.
\label{bdual}
\eeq
In 3d gauge theory, this is a geometrically corresponds to an exchange of the 2
$U(1)$ isometries (exchanging $(x_1, x_2)$ with $(x_3, x_4)$ in
\eqref{squashedmetric}).
This will appear as the modular duality of Liouville theory \cite{FKV1}, or more
mathematically as an identity
\eqref{sbselfdual} of quantum dilogarithm \cite{Faddeev95}, or as a modular 
double of a quantum group \cite{FaddeevModular}. 
This should be interpreted as an S-duality of the $SL(2,\bR)$
Chern-Simons
theory. We will make further comments on this in the last section.

\section{Basic Idea}\label{sec.idea}

\begin{figure}[htbp]
\centering{\includegraphics[scale=0.4]{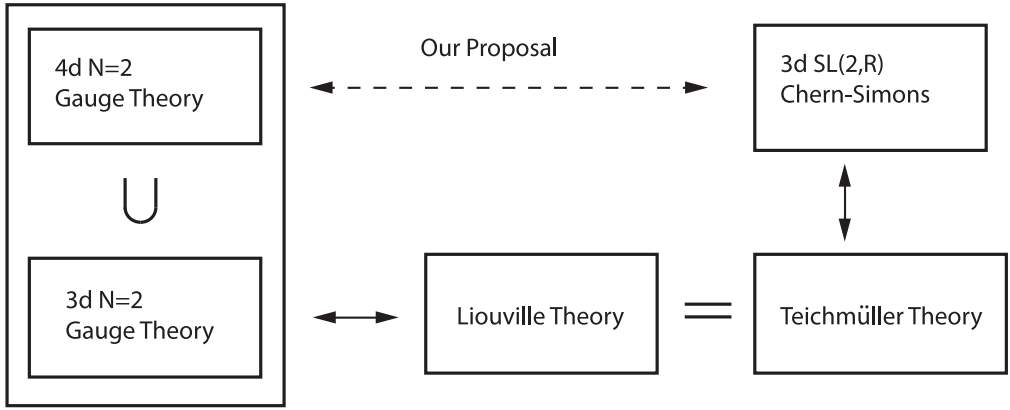}}
\caption{The flow of logic of this section (and the paper).}
\label{flow}
\end{figure}

In this subsection we give an evidence for the proposal above. In short, we
use quantum Liouville theory and quantum \Teichmuller theory as a bridge
between the two 3d theories (see Figure \ref{flow}). 
This is by no means a complete proof of our proposal, since we need to invoke
several conjectures existing in the literature, including the AGT
relation, which are assumed in this paper. Nevertheless the following
argument is a good starting point for the complete proof of our
proposal, and moreover provides a coherent perspective 
unifying all the theories discussed in this paper. Of course, the
physical reason for such a unification should be explained from the
existence of the mysterious 6d $(2,0)$ theory.

In the following we explain each of the steps in Figure \ref{flow} in some
more detail.

\subsection{Step 1: Gauge to Liouville}

Let us begin with the right hand side of \eqref{main1}, \eqref{main2}.
As we explained, the daughter 3d $\scN=2$ is defined as a theory on the duality
domain wall inside the mother 4d $\scN=2$ superconformal field theory, so let us begin
with this mother theory. When this
theory is placed on a compact manifold $S^4$, its
partition function can be computed by localization \cite{Pestun:2007rz},
and is given by
\beq
Z_{\rm 4d}[S^4](q;m;\epsilon_1,\epsilon_2)=\int \! da \,\, \nu(a)\,  \overline{Z_{\rm Nek}(q;a,m;\epsilon_1,\epsilon_2)} Z_{\rm Nek}(q;a,m;\epsilon_1,\epsilon_2),
\label{Pestun4d}
\eeq
where $a$ is the Coulomb branch parameter, $m$ mass parameters,
$q:=e^{2\pi i \tau}$ and $Z_{\rm Nek}$ is the instanton partition function studied by Nekrasov \cite{Nekrasov:2002qd},
with $\epsilon_1,\epsilon_2$ being the parameters of
$\Omega$-deformation. These two parameters can be traded for the two
parameters $b$ and $\hbar$, defined by
\beq
\epsilon_1= \hbar b , \quad \epsilon_2= \hbar /b.
\label{Omegaparameters}
\eeq
The overall parameter $\hbar$ is identified with the inverse radius of
$S^4$, and in the following we are going to set $\hbar=1$. Also, in
Pestun's computation we actually have $b=1$, although it is natural
to conjecture and existence of a 1-parameter family of manifold $S^4_b$ which reproduce
\eqref{Pestun4d} with $b\ne 1$.
In the following we are
going to deal with the right hand side of \eqref{Pestun4d} with $b\ne 1$. 
In the computation of localization, the two factors $Z_{\rm Nek}$ and
$\overline{Z}_{\rm Nek}$ are the contributions from the fixed points at
the north and the south pole, respectively, and the measure $\nu(a)$ contains classical as well as one-loop contributions.

AGT relation \cite{Alday:2009aq} states the equivalence of this partition function with a correlator of the Liouville theory on $\Sigma$. To fix notations, let us 
remind ourselves that Liouville theory is defined by an action
\beq
S=\int d^2 z \left(\partial \phi \bar{\partial} \phi + \pi \mu e^{2b\phi} \right).
\label{LiouvilleAction}
\eeq
This theory has central charge $c=1+6 Q^2$, where we defined
$Q=b+b^{-1}$. 
The primary operators are given by $V_{\alpha}=e^{\alpha
\phi}$, which has conformal dimension 
\beq
\Delta(\alpha)=\alpha (Q-\alpha).
\label{Deltaalpha}\eeq
Let us consider a correlator of these vertex operators.
When we specify the pants decomposition $\sigma$ of the Riemann surface, the correlator factorizes into the product of holomorphic and
anti-holomorphic conformal blocks $\scF^{\sigma}_{\alpha,E}$
\beq
\Big\langle \prod_i V_{m_i}  \Big\rangle_{\Sigma}=\int d\alpha \,
\nu(\alpha) \,  \overline{\scF_{\alpha,E}({q})}\scF_{\alpha,E}(q),
\label{Liouvillecorrelator}
\eeq
where $\alpha$ and $E$ denotes (the set of) internal and external
momenta, respectively, 
and $q$ ($\bar{q}$) denotes the complex structure (and its conjugate) of $\Sigma$.
We include the DOZZ 3-point function
\cite{Dorn:1994xn,Zamolodchikov:1995aa} 
in the measure $\nu(\alpha)$, which takes the form
\beq
\nu(\alpha)=\prod_i \sin (\pi \alpha_i b) \sin (\pi \alpha_i/b).
\label{nudef}
\eeq

Now the first claim of AGT is that Nekrasov partition function of the 4d
$\scN=2$ theory coincides
with the conformal block of Liouville theory, under the parameter
identification
\beq
\alpha=\frac{Q}{2}+a, \quad E=m.
\eeq
The second claim is that
the measure $\nu(a)$ from the classical and 1-loop partition function coincides with the product of DOZZ 3-point functions.
These yield the celebrated AGT relation, which claims the equivalence of a correlator with the 4d partition function
\beq
Z_{\rm 4d}[S^4]=\Big\langle \prod_i V_{m_i} \Big\rangle_{\Sigma}.
\eeq

4d gauge theories contain a rich class of BPS defect operators. For
example, Pestun's localization applies to the case with
a Wilson line operator. In general, we can consider the VEV of a 1/2 BPS loop
operator $L$ in 4d $\scN=2$ theory. 
The counterpart of this in Liouville theory is the Verlinde loop
operator $\scL$ \cite{Verlinde:1988sn}. There is a one-to-one correspondence between a
classification of charges of loop operators and the Dehn-Thurston data
of the non-self-intersecting curves on the Riemann surface
\cite{Drukker:2009tz}. Moreover, as shown in
\cite{Alday:2009fs,Drukker:2009id}, the VEV of the line operator in spin $j$
representation is equal to the Liouville correlator with a Liouville
loop operator $\scH$ corresponding to a degenerate field $\Phi_{1,2j+1}$ inserted:
\beq
\Big\langle L \Big\rangle_{S^4}=\Bigg\langle  \left(\prod_i V_{m_i}\right) \scL
\Bigg\rangle_{\Sigma}.
\label{DGOTrel}
\eeq

\begin{figure}[htbp]
\centering{\includegraphics[scale=0.3]{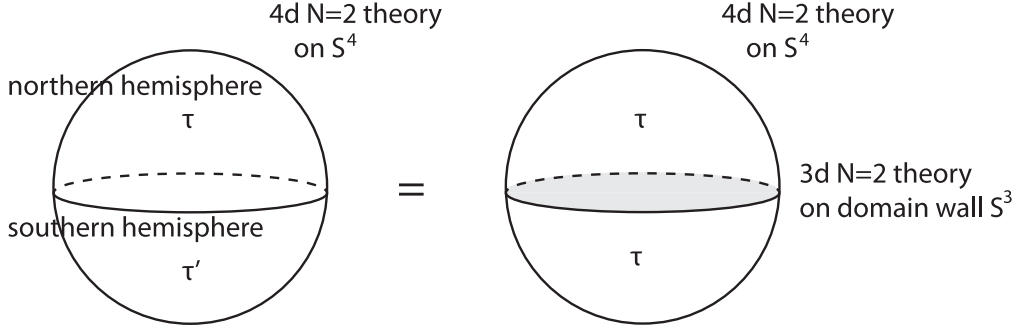}}
\caption{Our 3d $\scN=2$ theory is realized as a duality domain wall
 $S^3$ inside $S^4$, where the mother 4d $\scN=2$ theory lives. On the
 left figure, the complexified gauge coupling of the mother 4d theory in the northern
 (southern) hemisphere is given by $\tau$ ($\tau'$), where $\tau'$ is
 related to $\tau$ by an action of the S-duality group element $\varphi$:
 $\tau'=\varphi(\tau)$. Instead, we can take S-duality in the southern
 hemisphere to make the complexified gauge coupling $\tau$ in the whole
 $S^4$, but then we have a non-trivial 3d theory on the equator $S^3$.}
\label{S3inS4}
\end{figure}

In this paper we consider yet another type of operators in 4d gauge theories:
the domain wall operators. As already discussed in the previous subsection,
we are going to consider a 1/2 BPS domain wall. We introduced this as a
domain wall $\bR^3$ in $\bR^4$, but in order to make contact with
the story in this subsection we here consider a 1/2 BPS duality domain
wall placed on equator $S^3$ of $S^4$.
Summarizing, we have 4d theory on $S^4$, together with a duality wall on
the equator, and our 3d $\scN=2$ theory lives on this equator $S^3$
(Figure \ref{S3inS4}).

The domain wall preservers the symmetry used for
localization, and hence the partition function of the 4d theory is again
computed by localization. Precise computation of the 1-loop determinant can be
subtle, but \cite{Drukker:2010jp} proposed the following expression
\beq
Z_{\rm 4d+3d}(S^4)=\int da da' \,\nu(a) \nu(a')\, \overline{Z_{\rm Nek}(a,m)} Z_{\rm 3d}(a,a',m) Z_{\rm Nek}(a',m),
\label{Z4d3d}
\eeq
where the measure $\nu(a)$ is the same expression as before.
$Z_{\rm 3d}(a,a',m)$ is the partition function of 3d theory
$T[SU(2);\varphi;m]$ on $S^3$, with $a$ and $a'$
being either the mass parameter or the FI parameter of the 3d theory.
The 3d theory couples to the bulk by gauging the global symmetry.
$Z_{\rm Nek}$ and $\overline{Z_{\rm Nek}}$ are contributions from the
fixed points at the north pole and the south pole, respectively, and 
$Z_{\rm 3d}[a, a']$ is the contribution from the domain wall.

Now the natural question is whether there is a counterpart of this story
in the Liouville side. This has been analyzed in \cite{Drukker:2010jp},
and the answer is that it corresponds to an insertion of a
non-degenerate field, which changes \eqref{Liouvillecorrelator} into
\beq
\int d\alpha \, \nu(\alpha) \, \overline{\scF_{\alpha,E}(q)} 
\scF_{\alpha,E}(q'),
\label{FF}
\eeq
where $q'=e^{2\pi i \tau'}$ with $\tau'=\varphi(\tau)$.
Now the action of an element $\varphi$ of the mapping class on the conformal
block can be represented by an integral kernel
\beq
\scF_{\alpha,E}(q')=\int d\alpha' \nu(\alpha') \,
(\mathsf{\varphi})_{\alpha,\alpha'} \, \scF_{\alpha',E}(q).
\label{Ftransform}
\eeq
This is an analog of the modular transformation of characters in
rational conformal field theory (CFT) --- here we have an integral instead of a sum since our
CFT, Liouville, is irrational. For torus without punctures the kernel $\mathsf{\varphi}$ is indeed determined from the modular transformation of characters.

Substituting \eqref{Ftransform} into \eqref{FF}, we have
\beq
\int d\alpha d\alpha' \, \nu(\alpha) \nu(\alpha')\, \scF_{\alpha,E}(q) \varphi_{\alpha, \alpha'} \scF_{\alpha,E}(q) .
\label{Z2d3d}
\eeq
Comparing \eqref{Z4d3d} and \eqref{Z2d3d}, we conclude that the two
expressions are the same if we identify the partition function of the 3d
domain wall theory with an integral kernel representing $\varphi$
\cite{Drukker:2010jp,Hosomichi:2010vh}:
\beq
Z_{3d}[a,a']=\varphi_{\alpha, \alpha'}.
\label{Z=varphi}
\eeq
where the parameter identifications are given by
\beq
a=\alpha, \quad a'=\alpha'.
\eeq
In section \ref{sec.example}, we review existing checks and
provide further evidence for this conjecture \eqref{Z=varphi}. The relation
\eqref{Z=varphi} will be assumed for the discussion of the rest of this paper. 
and we call the equivalence of \eqref{Z4d3d} and \eqref{Z2d3d}
the ``generalized AGT relation''.

\subsection{Step 2: Liouville to \Teichmuller}

Now that we obtained the expression in Liouville theory, the next step
is to rewrite everything in \Teichmuller theory.

To explain why this is possible, let us begin with the classical theory.
The uniformization theorem states that for each element of the
\Teichmuller space 
there exists a unique constant negative curvature metric of the form
\beq
ds^2=e^{2 b \phi} dz d\bar{z}.
\label{2dmetric}
\eeq
This metric  has a constant negative curvature 
iff $\phi$ satisfies the Liouville equation
\beq
\partial \bar{\partial} \phi=2\pi \mu b e^{2b \phi},
\eeq
which coincides with the equation of motion for the Liouville action
\eqref{LiouvilleAction}. This is the classical equivalence between Liouville and \Teichmuller theory.

There is a quantum counterpart of this equivalence.
When we write the two Hilbert spaces for the two
theories on a Riemann surface $\Sigma$ by $\scH_L(\Sigma), \scH_T(\Sigma)$,
we have
\beq
\scH_L(\Sigma)=\scH_T(\Sigma),
\eeq
where the equality is meant to include the equality of the mapping class
group action as well.
This was originally conjectured by a seminal paper by
H.~Verlinde \cite{Verlinde:1989ua} in the 80's. 
The explicit construction of quantum
\Teichmuller theory was given in the mid 90's
\cite{Chekhov:1999tn,KashaevQuantization}, and more recently
there are important contributions \cite{Teschner:2003at,Teschner:2005bz}
which almost proves the equivalence of
the two theories. See also \cite{FKV2}, which established a direct
correspondence between quantum discrete Liouville theory and quantum
\Teichmuller theory.

Quantum \Teichmuller theory is a framework to construct the Hilbert
space $\scH_T(\Sigma)$. The precise definition of the Hilbert space of quantum \Teichmuller
space will be given in appendix \ref{sec.Teichmuller}, and we will give
concrete discussion in section \ref{sec.example}. Here we summarize
the minimal ingredients needed for the purpose of this section.

It is important for our purposes that there are several natural different bases in $\scH_T(\Sigma)$.
One basis is the holomorphic basis $|\tau \rangle$, where $\tau$ is a 
holomorphic coordinate parametrizing the complex structure of $\Sigma$.
This arises from
the \Kahler quantization of the \Teichmuller space.

Another is the length basis $|l \rangle$.
This is obtained when we
specify a pants decomposition $\sigma$, and
where $l=\{ l_i \}$ are eigenvalues of the geodesic length operators and takes
values in $\bR_{>0}$. 
In the language of Liouville theory, these length operators are
precisely the Verlinde
loop operators \cite{Drukker:2009id} ($\scL$ in \eqref{DGOTrel}), and 
the basis $|l\rangle$ is the eigenspace of maximally commuting set of
Verlinde loop operators determined by the pants decomposition.
The important property of the basis 
is that they span a complete basis in $\scH_{\Sigma}$ \cite{KashaevQuantum}
\beq
\langle l | l' \rangle =\nu(l)^{-1} \delta(l-l'),  \quad \int dl \,\nu(l)\, |l \rangle \langle l|=1,
\label{nucomplete}
\eeq
where $\nu(l)$ is the same function as defined in \eqref{nudef}.

What we would like to do from now on is to rewrite the expressions
\eqref{Liouvillecorrelator}, \eqref{Z2d3d}
in the language of quantum \Teichmuller theory.
The crucial observation for this is that
the conformal block of the Liouville
theory can be identified with the overlap of holomorphic basis and
length basis in quantum \Teichmuller theory \cite{Teschner:2003at}:
\beq
\scF_{\alpha,E}(q)=\langle l| q \rangle,
\label{FinTeichmuller}
\eeq
where the length parameters $l$ are identified with the Liouville
momentum $\alpha$
\beq
\alpha=l,
\label{alpha=l}
\eeq
and the puncture parameters $m$, corresponding to external momenta $E$, are
suppressed in the notation of the right hand side.
For consistency of this equation, note that both sides depend on the
choice of the pants decomposition (which is suppressed in the notation
above), as well as on mass parameters.
As explained in 
\cite{Teschner:2010je}, this follows
since both sides (1) transform the same way under the action of the
mapping class group and (2) has the same asymptotic behavior. See
\cite{Teschner:2001rv,Teschner:2003en} for computations on the Liouville side.

The parameter identification \eqref{alpha=l} means that conformal blocks
are labeled by length $l$. 
This has been anticipated long ago, see
\cite{Verlinde:1989ua,Witten:1990wn}. We also review one supporting
argument for this in appendix \ref{subsec.length}. The parameter $l$ is an analogue of discrete
labels of the 
conformal blocks in rational CFT. It is not surprising that the label
$l$ now becomes a continuous parameter, specifying a continuous
representation of $SL(2,\bR)$. What is surprising here is that $l$ is at
the same time the label for the complex structure moduli; in rational
CFT's, the discrete labels of the conformal blocks and the complex
structure moduli (continuous parameter) are different variables, whereas
here the two are unified into a single continuous parameter $l$.

The parameter identification \eqref{alpha=l} immediately means that we should have
\beq
Z_{3d}(a,a')=\varphi_{l,l'}=\langle l| \mathcal{\varphi}| l'\rangle, 
\label{lvarphil} 
\eeq
where we used the same symbol $\varphi$ for the operator
in the Hilbert space of quantum \Teichmuller theory.
In other words, the modular kernel is simply the matrix representation
of $\varphi$ in the basis $|l\rangle$. This is the quantum \Teichmuller
version of the 3d partition function in \eqref{main1}.
This again depends on the choice of the pants decomposition, which 
on the gauge theory side determine the Lagrangian description of the
mother 4d theory.

To see the consistency of this relation, let us start with
\beq
\varphi |q\rangle =|q' \rangle, \quad
\eeq
where $q'=e^{2\pi i \varphi(\tau)}$.
These equations will follow from the definition of the state $|q\rangle$
but its meaning is intuitively obvious.
We therefore have
\begin{equation}
\begin{split}
\scF_l(q')&=\langle l| q' \rangle
= \langle l| \varphi | q \rangle \\
&= \int \!dl' \, \nu(l') \, \langle l| \varphi | l'\rangle \langle l'| q \rangle
= \int \! dl' \, \nu(l') \, \varphi_{l,l'}\scF_{l'}(q).
\label{F=Fderiv}
\end{split}
\end{equation}
This reproduces the transformation property \eqref{Ftransform} of the conformal block.

For the convenience of the reader in Table \ref{triality} we have summarized the
correspondence between gauge/Liouville/\Teichmuller theories discussed
so far.

\begin{table}[htbp]
\caption{Dictionary relating gauge theory, Liouville and \Teichmuller theory.}
\begin{center}
\begin{tabular}{|c|c|c|}
\hline
4d/3d gauge theory & Liouville & \Teichmuller \\
\hline
\hline
S-duality & mapping class group & mapping class group \\
\hline
mass/FI parameter $a, a'$ & internal momenta $\alpha, \alpha'$ & length parameters
	 $l,l'$ \\
\hline
mass parameter $m$ & external momentum $E$ & puncture parameter $m$ \\
\hline
line operator $L$ & Verlinde loop operator $\scL$ & 
geodesic length operator $\scL$\\
\hline
3d partition function  & integral kernel for $\varphi$
    & VEV of operator $\varphi$ \\
$Z_{\rm 3d}(a,a')$  &  $\varphi_{\alpha,\alpha'}$ & $\langle l | \varphi
	 | l'\rangle$\\
\hline
Nekrasov partition function & conformal block &
	 pairing \\
$Z_{\rm Nek}(a,m)$ & $\scF_{\alpha,E}$ & $\langle q | l \rangle$ \\
\hline
\end{tabular}
\end{center}
\label{triality}
\end{table}

\subsection{Step 3: \Teichmuller to Chern-Simons} \label{subsec.Step3}

So far everything is defined in two dimensions, but we would like to
lift this story to three dimensions: we will find a $SL(2,\bR)$ Chern-Simons theory.

Let us again start with the classical theory.
We write the 2d metric on $\Sigma$ by zweibeins $e^+, e^-$
\beq
ds^2=e^+\otimes e^-,
\eeq
and by the  $SO(2)$ spin connection $\omega$. When we regard $\omega$ as
an independent degrees of freedom, we have
\beq
\scT_{g,h}= \frac{ \{ (e^+, e^-, \omega;\, \scG_a=0) \}
}{\mathrm{Diff}_0\times \mathrm{LL}}, 
\eeq
where LL denotes local Lorentz transformations. The constraints are given by
\begin{equation}
\begin{split}
\scG^+&=de^+-\omega\wedge e^+ , \\
\scG^-&=de^-+\omega \wedge e^- , \\
\scG^0&=d\omega -e^+\wedge e^-.
\label{constraintG}
\end{split}
\end{equation}
The first two equations are the definitions of the spin connection,
whereas the third condition is the condition of constant negative
curvature.

Let us define
\beq
\scA=e^+ T^{-} +e^- T^+ +\omega T^3,
\eeq
where $T^+, T^-, T^3$ are generators of $SL(2,\bR)$, satisfying commutation relations
\beq
[T^3, T^+]=T^+, \quad [T^3, T^-]= -T^-, \quad [T^+, T^-]=T^3.
\eeq
Then the constraints \eqref{constraintG} are translated into the condition that 
$\scA$ is a $SL(2,\bR)$ flat connection:
\beq
\scF=d \scA+\scA\wedge \scA=0.
\label{flatF}
\eeq
We recognize this as the equation of motion of the $SL(2,\bR)$ Chern-Simons
theory. At the level of the Lagrangian we see that the action of the
$SL(2,\bR)$ Chern-Simons theory, when written in terms of the zweibein and
the spin connection, takes the form
\begin{equation}
\begin{split}
S[e^+,e^-,\omega]&=\frac{k}{4\pi} \int_{\bR\times \Sigma}
 \left(\frac{1}{2}\omega \wedge d\omega+ e^+ \wedge de^- +\omega\wedge e^+ \wedge e^- \right) \\
&=\frac{k}{4\pi} \int_{\Sigma} \left(\frac{1}{2}\omega\wedge \partial_t
 \omega+e^+\wedge \partial_t e^- +\omega_t \wedge \scG^0+e^-_t \wedge \scG^{+} +e^+_t \wedge\scG^-  \right),
\label{Seomega}
\end{split}
\end{equation}
In the last expression we integrated over the time variable $t$, and 
we recover the constraints \eqref{constraintG}.
It is remarkable that we have a manifest (2+1)-dimensional invariance,
and it is one of the important purposes of this paper 
to recover this (2+1)-dimensional invariance for our partition
functions.

\bigskip

There are some caveats in the classical 2d metrics and the $SL(2,\bR)$
Chern-Simons theory mentioned above, and it is important to keep this subtlety in mind.
Not all classical solutions of
$SL(2,\bR)$ Chern-Simons have their counterparts in \Teichmuller
theory. For example, $\scA=0$, i.e. $e^+=e^-=\omega=0$ is a classical
solution of \eqref{flatF}, but the corresponding metric is trivial and
is not non-degenerate. Such a singular metric is not allowed in gravity, or 
at least in classical gravity. 
In general, flat connections are classified by Wilson loops, i.e.
\beq
{\rm Hom}\left(\pi_1(\Sigma), SL(2,\bR)\right)/SL(2,\bR),
\eeq
where the gauge symmetry $SL(2,\bR)$ on the right acts by conjugation. It is known that 
this space has several connected components\footnote{More precisely, the
mathematical statements which follow in this paragraph is for
$PSL(2,\bR)$ flat connections. The relevant connections, however, can be
lifted to $SL(2,\bR)$.}. Each gauge field (a connection) specifies a vector bundle, 
and its Euler number labels the connected components. For a flat bundle,
the absolute value of this 
number is bounded by $2g-2$ (\cite{Milnor,Wood}, see also
\cite{HitchinSelfDuality}), and the \Teichmuller space is the identified
with the component with the maximal value $2g-2$ \footnote{The Euler number
changes sign when we change the orientation of $\Sigma$. When we take
this into account, we essentially have $2g-1$ connected components.}.
This is an important difference between Chern-Simons theory and
\Teichmuller theory. 
In the following we restrict ourselves to discussion involving only the local structure of the moduli space
of $SL(2,\bR)$ flat connections in the \Teichmuller component, and 
will neglect more global structures of the moduli space\footnote{It
is probably the case that our
theory, where the metric is required to be invertible, is actually
closer to (2+1)-dimensional gravity than Chern-Simons
theory.}. 

\bigskip

We now want to consider 3d $SL(2,\bR)$ Chern-Simons theory with
boundary. The key result crucial for the discussion here is the
following: when we have a $SL(2,\bR)$ Chern-Simons theory on a 3-manifold with boundary,
 on the boundary we have a Liouville theory
 \cite{Verlinde:1989ua,Verlinde:1989hv,Carlip:1991zm}.

This should be considered as an analogue of the classic correspondence between $SU(2)$
Chern-Simons theory and Wess-Zumino-Novikov-Witten (WZNW) model
\cite{Witten:1988hf,Elitzur:1989nr}. Some readers, however, might be
puzzled by this statement, since it is also stated in the literature 
that $SL(2,\bR)$ Chern-Simons theory on a 3-manifold with a boundary has
$SL(2,\bR)$ WZNW model on the boundary. In fact, the difference between
these two statements arise from the choice of the boundary
conditions. This is explained clearly in \cite{Verlinde:1989ua}, so let
us briefly summarize the results there\footnote{We would like to thank
H.~Verlinde for explanation of his work.}.

\bigskip

Consider $SL(2,\bR)$ Chern-Simons theory on a manifold with boundary.
The action \eqref{SL2Raction} is then not invariant under the 
infinitesimal change of $\scA$, and 
we need to supplement the action with a boundary condition
to make the action invariant.
One way to specify a boundary condition is to set the values of the fields
to be constant at the boundary. 
For this purpose we need to choose a polarization ---
we divide the phase space degrees of freedom into coordinates and their
conjugates, the momenta. In $SL(2,\bR)$ Chern-Simons theory a natural choice is
a holomorphic polarization, where we take $A_z^a$ ($a=1,2,3$) as coordinates.
Here $z$ and $\bar{z}$ represents the coordinates on the 2d surface on
which we do canonical quantization.
The conjugate variable $A_{\bar{z}}^a$ then becomes
\beq
A^a_{\bar{z}}=i \frac{4\pi}{k}\frac{\partial}{\partial A_z^a}\,\,.
\eeq
The well-known results in Chern-Simons theory says that after solving the Gauss law constraints (quantum version of
\eqref{flatF} imposed on the wavefunction), we have a wave function
\beq
\Psi[A^0_z,A^+_z,A^-_z]=\exp \left(\frac{ik}{4\pi}S_{\rm WZNW}\right),
\eeq
where $\Gamma$ is the $SL(2,\bR)$ WZNW action
\beq
S_{\rm WZNW}=\int_{\Sigma} \textrm{Tr}\left(g^{-1}\partial g\right) (g^{-1}
\bar{\partial} g)+\int_M \textrm{Tr}( g^{-1} d g)^3,
\quad A_z^{0,\pm}=(g^{-1}\partial g)^{0,\pm},
\eeq
where $\Sigma$ is the boundary of $M$. This is the correspondence
between WZNW model and the Chern-Simons theory well-known in the
literature.

We can also choose a different polarization. As discussed in subsection
\ref{subsec.Step3}, we can trade $\scA$ for the zweibein $e^+, e^-$
and the spin connection $\omega$. We choose $\{\omega_z, e^+_z, e^{+}_{\bar{z}}\}$ as independent degrees of freedom\footnote{This
polarization is not independent of $SL(2,\bR)$ gauge transformation.}. The conjugate
variables become (see \eqref{Seomega})
\beq
\omega_{\bar{z}}=i \frac{4\pi}{k} \frac{\partial}{\partial \omega_z}, \quad
e_{z}^-=i \frac{4\pi}{k} \frac{\partial}{\partial e^+_{\bar{z}}}, \quad
e_{\bar{z}}^-=-i \frac{4\pi}{k} \frac{\partial}{\partial e^+_z}\,\,
.
\eeq
The constraint equations
\eqref{constraintG}, imposed on the wavefunction 
$\tilde{\Psi}[\omega_z, e^+_z, e^+_{\bar{z}}]$,
are now non-linear, but fortunately we can solve them.
When we write
\beq
e^+= e^{\phi}(dz+\mu d\bar{z}), \quad 
e^-= e^{\bar{\phi}}(d\bar{z}+ \bar{\mu} dz),
\eeq
we have a wavefunction $\tilde{\Psi}[\omega_z, \phi, \mu]$. 
This wavefunction itself is a coordinate dependent quantity since
$\omega_z$ manifestly depends on the
choice of the coordinate $z$. However, this dependence drop out when
we form an inner product of wavefunctions and carry out a Gaussian
integral with respect to $\omega_z$; we are then left with integrals over
$\phi$ and $\mu$.
The $\phi$ dependence of the wavefunction is a chiral half of the
Liouville action, and when we fix a gauge by choosing a conformal gauge,
we have an inner product of the Liouville theory, integrated over the
Teichmuller space.
This is the derivation of the statement that 
in $SL(2,\bR)$ Chern-Simons theory there is a boundary condition labeled by the complex structure
moduli such that the boundary theory is a Liouville theory.

Of course, the two polarizations are not unrelated, and 
one can recover one from the other; the
wavefunction in the two polarizations are related by a Legendre
transformation, and we are able to obtain Liouville theory from 
WZNW model by a procedure known as Hamiltonian reduction or
Drinfeld-Sokolov 
reduction \cite{Alekseev:1988ce,Bershadsky:1989mf,Forgacs:1989ac}.
The existence of this hidden $SL(2,\bR)$ symmetry in Liouville
theory was discovered by the work of Polyakov \cite{Polyakov:1987zb}.
See also \cite{Coussaert:1995zp} for the discussion in terms
of asymptotic boundary conditions of 3d gravity.


\bigskip
In general, a useful framework for studying the connection between 
a topological 3d theory on a 3-manifold and another theory
on its 2d boundary is the axiomatic formulation of topological quantum
field theory (TQFT) \cite{AtiyahTopological}. 
Suppose that we have a 3d topological quantum field theory. When we have
a 3d manifold $M$ without boundary, we have a number, the partition
function on $M$. When we have a 2d surface $\Sigma$, we have a Hilbert
space $\scH(\Sigma)$ on it, since we can canonically quantize the theory
on $\Sigma\times \bR$, where $\bR$ direction is regarded as time.
When we have a 3-manifold $M$ with boundary $\Sigma$, we can do a path
integral over $M$, and we have an element $|M\rangle $ of
$\scH(\Sigma)$. When 
$M$ has two boundaries ($\partial M=(-\Sigma)\cup \Sigma'$, minus sign
representing the orientation reversal), path integral
over $M$ gives a map from $\scH(\Sigma)$ to $\scH(\Sigma')$.

\begin{figure}[htbp]
\centering{\includegraphics[scale=0.3]{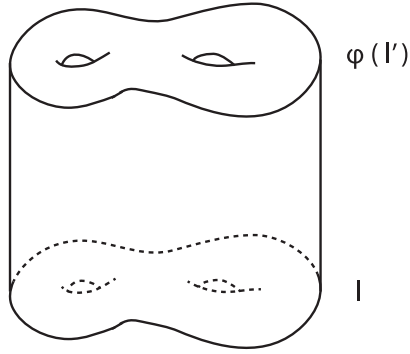}}
\caption{We consider Chern-Simon theory on $\Sigma\times I$, where the
 complex structure of the two Riemann surfaces on the boundary are
 specified by $l$ and $\varphi(l')$.}
\label{cobordism}
\end{figure}

Let us apply this general framework to the Hilbert space of the quantum
\Teichmuller theory. We are then lead to the identification of the
Hilbert space of the Chern-Simons theory with that of the quantum
\Teichmuller theory:
\beq
\scH_T(\Sigma)=\scH_{\rm CS}(\Sigma),
\eeq
where we emphasize again that on the right hand side we have specified
the boundary condition on $\Sigma$ by an element of the \Teichmuller space.
In particular, this includes the statement that the
action of the mapping class group commutes with the isomorphism between
the Hilbert spaces. This means that the partition function of the 3d
domain wall theory, which we now know is equivalent to
$\langle l| \varphi | l'\rangle$ \eqref{lvarphil}, can be understood as a transition
amplitude in Chern-Simons theory: the partition function on a 
3-manifold $M=\Sigma\times I$, where the boundary conditions on 
$\Sigma\times \{0\}$
and $\Sigma\times \{1\}$ are determined by $l$ and $\varphi(l')$ (see
Figure \ref{cobordism}).
This gives the right hand of the relation \eqref{main1}:
\beq
Z_{\rm CS}[\Sigma\times I]=\langle l | \varphi | l' \rangle.
\eeq
We can also take a sum over $l$, which geometrically corresponds to
replacing $\Sigma\times I$ by $\Sigma\times S^1$:
\beq
Z_{\rm CS}[\Sigma\times S^1]=\textrm{Tr}(\varphi)=\int\! dl\, \langle l | \varphi | l'\rangle.
\label{tracevarphi}
\eeq
This is the right hand side of \eqref{main2}.

\subsection{\texorpdfstring{$SL(2,\bR)$}{SL(2,R)} Chern-Simons Reformulation of the Generalized AGT relation}
\label{subsec.reformulation}

In the discussion so far our aim has been to provide evidence for
\eqref{main1} and \eqref{main2}. However,
we have already obtained more than we originally aimed for; all the
ingredients of the generalized AGT relation \eqref{Z4d3d} can now be
expressed in the language of the \Teichmuller theory, and consequently
the $SL(2, \bR)$ Chern-Simons theory. This give a natural question:
is there a natural interpretation of the whole expression \eqref{Z4d3d},
not just the 3d domain wall partition function?

Using the expression \eqref{FinTeichmuller} for the conformal block in
quantum \Teichmuller theory, \eqref{Z2d3d} becomes
\beq
 \int dl dl' \, \nu(l) \nu(l')\, \langle \bar{q} |l \rangle \langle l| \varphi | l'\rangle \langle l'|q \rangle.
\eeq
Using the completeness relation of basis $|l\rangle$
 \eqref{nucomplete}, this reduces to a rather simple expression
\beq
\langle \bar{q} |\varphi| q \rangle=\langle \bar{q} | q' \rangle.
\label{tautau}
\eeq
Again, we have an expectation value of the operator $\varphi$, but now
 in the holomorphic basis. This is to be expected since conformal block
 represent the change of basis in quantum \Teichmuller theory.
\eqref{tautau} represents the two figures in Figure \ref{S3inS4}; for
 example, the right hand side simply says that we have a Janus solution with $\tau$ in the northern
 hemisphere and $\tau'$ in the souther hemisphere,
which is represented by a pairing
 between the states $|q\rangle$ and $|q' \rangle$ in the
Hilbert space $\scH_{\rm CS}(\Sigma)$ (recall $q:=e^{2\pi i \tau}$).

The discussion so far raises the following question: is there a
counterpart of the state $|q\rangle$ in 3d domain wall theory, not
just its pairing? Is there a Hilbert space in 4d theory?
Since the 4d theory near the domain wall is defined on
$S^3\times I$, by regarding $I$ direction as time and canonically
quantizing the theory we have a natural candidate,
the Hilbert space of quantum
ground states of the 4d theory on $S^3$. We are going to denote this by
$\scH_{\rm 4d}(S^3)$. From this viewpoint the chiral half of the AGT relation can be formulated
as an equivalence of two Hilbert spaces, $\scH_{\rm 4d}(S^3)$ and
$\scH_L(\Sigma)$,
where as we have seen the latter coincides with $\scH_{\rm CS}(\Sigma)$. Moreover,
the Nekrasov partition function and the conformal block are represented by
the same element in the Hilbert space in this identification:
\begin{equation}
\begin{split}
\scH_{\rm 4d}(S^3)&=\scH_{\rm L}(\Sigma), \\
\rotatebox{90}{$\in$}\hspace{5mm} & \hspace{5mm} \rotatebox{90}{$\in$}\\ 
|Z_{\rm Nek}\rangle &= |\scF \rangle,
\end{split}
\end{equation}
and the equivalence of the 4d partition function \eqref{Z4d3d} and
\eqref{Z2d3d} becomes a simple statement that an action of the matrix element of
$\varphi$ commutes with the isomorphism between the Hilbert spaces of the two theories:
\beq
\langle \overline{Z_{\rm Nek}} | \varphi |Z_{\rm Nek} \rangle=
\langle \overline{\scF} | \varphi |\scF \rangle.
\label{simple}
\eeq

The interpretation of the AGT correspondence as an identification of
$\scH_{\rm 4d}(S^3)$ and $\scH_{L}(\Sigma)$ is not new, see for example
\cite{Drukker:2009id,Nekrasov:2010ka}. The novelty of the 
present discussion is that we consider an action of the mapping class
group and gave a Chern-Simons interpretation of the results.

There is actually one more Hilbert space in the literature.
In \cite{Nekrasov:2010ka}, Nekrasov and Witten considered a 4d topologically twisted theory
on $\Omega$-deformed 3-sphere, $S^3_{\epsilon_1, \epsilon_2}$,
where $\epsilon_1, \epsilon_2$ are parameters of $\Omega$-deformation \eqref{Omegaparameters}.
Pestun's result then means that
quantum ground state of the physical string theory on $S^3\times \bR$
coincides with the Hilbert space of the topological twisted theory on
$\Omega$-deformed $S^3_{\epsilon_1, \epsilon_2}$, with $\epsilon_1=\epsilon_2$.
It is natural to extend this result to $b\ne 1$;
quantum ground state of the physical string theory on $S^3_b$
should coincide with the Hilbert space of the topological twisted theory on
$\Omega$-deformed $S^3_{\epsilon_1, \epsilon_2}$:
\beq
\scH_\textrm{topological 4d}(S^3_{\epsilon_1, \epsilon_2})
=
\scH_\textrm{physical 4d}(S^3_b).
\label{topologicalphysical}
\eeq
It is an interesting problem to find out if this is correct. Note that
 both $S^3_b$ and $S^3_{\epsilon_1, \epsilon_2}$
preserve the same symmetry $U(1)\times U(1)$.
If \eqref{topologicalphysical} is correct, then 
by taking the limit $\epsilon\to 0$, our story should be directly
related with the work of Nekrasov and Shatashvili \cite{Nekrasov:2009rc}.
Works are currently in progress in this direction.


\bigskip
As already explained, the simple structure of the left hand side in
\eqref{simple} 
has a geometric meaning: $\langle \bar{Z}|$ corresponds to a contribution from a
northern hemisphere, $\varphi$ to a domain wall on $S^3$, and
$|Z\rangle$ to a southern hemisphere. In other words, we have a
decomposition of the 4-sphere
\beq
S^4=D^4\cup_{S^3} D^4,
\eeq
where $D^4$ is a disc (hemisphere) which has the equator $S^3$ as a boundary.

The counterpart of this decomposition in the $SL(2,\bR)$ Chern-Simons theory
will be as follows. Choose a 3-manifold $M_{\tau}$ with boundary
$\Sigma$ with complex structure $\tau$, such that the path integral of
the Chern-Simons theory gives an element $|q\rangle \in \scH(\Sigma)$. 
We then have a 3-manifold
\beq
M=M_{\tau} \cup_{\varphi} M'_{\tau'},
\label{Mdecomp}
\eeq
and
\beq
Z_{\rm 4d}(S^4)=Z_{\rm CS}(M).
\label{main3}
\eeq
It is notable that the relation \eqref{main3} claims an equivalence of the
partition function of {\it 4d} $\scN=2$ theory (with 3d 1/2 BPS defect) and 3d Chern-Simons theory!

Unfortunately, we have not been able to explicitly identify the manifold
$M_{\tau}$ which gives rise to a state $|q\rangle$ in $\scH(\Sigma)$.
However, there is an interesting observation: the decomposition
\eqref{Mdecomp}
is reminiscent of the Heegaard decomposition of 3-manifolds.

\begin{figure}[htbp]
\centering{\includegraphics[scale=0.4]{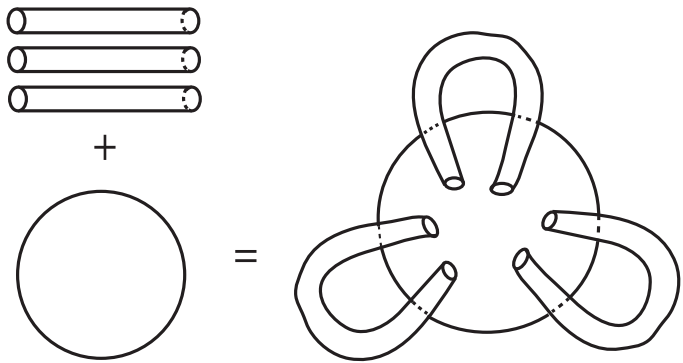}}
\caption{A genus $g$ handlebody is a 3-manifold obtained by attaching
 attaching $g$ handles to a 3-ball. Its boundary is a genus $g$ Riemann surface.}
\label{handlebody}
\end{figure}

To describe this, let us define a genus $g$ handlebody $H_g$ as a 
3-disc $D^3$ with $g$ handles attached to it (see Figure \ref{handlebody}). The boundary of this
3-manifold is a genus $g$ Riemann surface without punctures, $\Sigma_g$.
This means that when we have two handlebodies of the same genus $g$ and
an element $\varphi$ of the mapping class group of $\Sigma_g$, we can
glue two handlebodies to construct a closed 3-manifold.
\beq
M=H \cup_{\varphi} H',
\label{Heegaard}
\eeq
where $\varphi$ maps $\partial H$ to $\partial H'$. This decomposition is
called a Heegaard decomposition of $M$, $\Sigma$ is called a Heegaard
surface, and $g$ the Heegaard genus. There is a theorem saying that
every closed 3-manifold admits a Heegaard
decomposition\footnote{Simple examples: $S^3$ can be decomposed into northern and southern hemispheres
at the equator $S^2$, which gives a genus $0$ Heegaard decomposition. Lens spaces
have genus $1$ Heegaard decompositions.}.

Comparing the two expressions \eqref{Mdecomp} and \eqref{Heegaard}, it
is tempting to identify $M_{\tau}$ and $H$, when our Riemann surface
$\Sigma=\Sigma_{g,h}$ does not have a puncture, $h=0$. This is a
conjecture we make in this paper. If this turns out to be the case, 
$SL(2,\bR)$ partition function of an arbitrary 3-manifold, not just that
 on $\Sigma\times I$, has a direct interpretation in 4d $\scN=2$ gauge theory.
In fact, in the case of $SU(2)$ Chern-Simons theory, the same strategy
was taken to define an invariant of the 3-manifold \cite{KohnoTopology}.
The method of the paper \cite{KohnoTopology} applies directly to any
rational conformal field theories, but not to irrational theories as
discussed in this paper\footnote{Technically, we have to show that the
invariant defined does not depend on the choice of the Heegaard
splitting. 
}. It would be interesting to
investigate this point in more detail.

Our proposed correspondence is summarized in Table \ref{SummaryTable}.

\begin{table}[htbp]
\caption{Summary of the proposed correspondence between 4d/3d gauge
 theory and $SL(2,\bR)$ Chern-Simons theory.}
\begin{center}
\begin{tabular}{|c|c|}
\hline
4d/3d gauge theory & 3d $SL(2,\bR)$ Chern-Simons \\
\hline
\hline
$S^4$ with duality wall on $S^3$  & 3-manifold $M$ \\
\hline
decomposition of $S^4$ & Heegaard decomposition \\
$S^4$=$D^4 \cup _{S^3} D^4$  & $M=H_1 \cup_{\Sigma} H_2$ \\
\hline
disc $D^4$ with boundary $S^3$ & handlebody $H$ with boundary $\Sigma$\\
\hline
duality wall $S^3$ & mapping class group of $\Sigma$ \\
\hline
quantum ground space of gauge theory &  Hilbert space
     on the boundary\\
$\scH_{\rm 4d}(S^3_b)$ & $\scH_{\rm CS}(\Sigma)$\\
\hline
Nekrasov partition function $Z_{\rm Nek} \in \scH_{\rm 4d}(S^3_b)$ & 
conformal block $\scF \in \scH_{\rm CS}(\Sigma)$ \\
\hline
partition function on $S^4$ & partition function on $M$ \\
\hline
\end{tabular}
\end{center}
\label{SummaryTable}
\end{table}

\bigskip
\noindent {\it Comments on the M5-brane Interpretation} 

As already mentioned in introduction, the relation \eqref{main1} can be considered as a 3+3 analog of the AGT
relation,
and has an M5-brane interpretation. 
Recall that AGT correspondence arises from M5-branes on
$(S^3\times I)\times  \Sigma$,
where 4d $\scN=2$ theory lives on $S^3\times I$ and 2d Liouville theory
on $\Sigma$. In this language, our correspondence can be thought of as a
M5-brane 
theory on $S^3\times (I\times \Sigma)$.
In this sense our story can be considered as a (3+3) analog of AGT
relation, where one theory (3d $\scN=2$ domain wall theory) 
lives on $S^3$ and another ($SL(2,\bR)$ Chern-Simons theory)
on $I\times \Sigma$.
However, we should keep in mind the 
differences --- our story includes extra domain walls, which breaks the
supersymmetry to half ($\scN=2$ in 3d), as opposed to 4d $\scN=2$. In
M-theory this will be represented by 
the existence of another
M5-branes. The intersection of
two M5-branes will be codimension 2 defects in the original
M5-brane (cf.~\cite{Drukker:2010jp}).
Brane configurations reminiscent of ours appears in the recent work of
\cite{Witten:2011zz}, which includes (as a particular case of more
general brane setups) D3-branes wrapping $S^4$ and an NS5-brane on the
equator $S^3$ inside $S^4$.

Before closing this section, let us comment on (possibly) related proposals in the literature. For example,
\cite{Ganor:2008hd,Ganor:2010md} conjectures a Chern-Simons description
of 4d $\scN=4$ theory on $S^1$ with twists by R-symmetry and S-duality\footnote{Some people might be tempted to relate $S^3\times S^1$
version of the proposal \eqref{main2} with the one of
\cite{Gadde:2009kb}, which also consider the 4d theory on $S^3\times
S^1$. However, there are important differences. First, in that
reference they consider a superconformal index of the 4d theory, not the partition
function of the 4d theory with 3d defects. Second, their index is
invariant under the modular transformations --- this is why they have 2d
TQFT, not 3d TQFT.}. \cite{Dimofte:2010tz} proposes
a correspondence between 3d $SL(2,\bC)$ Chern-Simons theory and 2d
$\scN=(2,2)$ theory, which could presumably be understood as a dimensional
reduction of our story here. \cite{Wu:2009tq} discuss the Chern-Simons
reformulation of AGT relation, and in particular makes contact with the
results of \cite{Alday:2009fs,Drukker:2009id}.
\cite{Nekrasov:2011bc} also discusses the 3d
hyperbolic geometry in the context of AGT relation and the work of
\cite{Nekrasov:2009rc}, while the
2d Teichm\"uller/3d Chern-Simons connection as in section \ref{sec.CS}
has already appeared in \cite{DimofteTalk}. 
The novelty of the present paper is that we have provided 
coherent presentation of all the topics from a unified perspective,
starting from 4d/3d supersymmetric gauge theories to 3d $SL(2,\bR)$
Chern-Simons theory.
We also have clarified the meaning of the AGT relation, and have provided 
exact 
quantitative statements \eqref{main1}, \eqref{main2}, 
which can be checked by explicit computations.
It would be interesting to clarify the precise relation of our
proposals and those in the literature. Some works in this direction is currently
in progress \cite{InProgress}.

\section{Example: Once-Punctured Torus} \label{sec.example}

In this section we work out the
example of $\Sigma_{1,1}$ in detail\footnote{The AGT relation in this
example was proven in \cite{Fateev:2009aw}.}.
Many of the our results below can straightforwardly be adopted for more
general examples (although computationally more complicated), 
and we have included several arguments which are independent of 
the example we discuss here.

\subsection{Theories on Duality Walls}

Let us begin with the description of the mother 4d $\scN=2$ theory and
its daughter, the 3d $\scN=2$ domain wall theory.
For once-punctured torus, the mother 4d theory
is the 4d $\scN=2^*$ theory, where the parameter associated with the
puncture of $\Sigma_{1,1}$ plays the role of an adjoint mass parameter
deforming $\scN=4$ to $\scN=2^*$. The mapping class group of
$\Sigma_{1,1}$ is given by $SL(2,\bZ)$, which is identified with the S-duality group
of the mother 4d theory. 

What we would like to do is to identify the action of this
group
on the 3d $\scN=4$ theory. This $SL(2,\bZ)$ action is not a symmetry of the
3d theory itself; rather it maps one 3d theory to a different 3d theory.
This $SL(2,\bZ)$ action was studied by \cite{Witten:2003ya}
for an Abelian gauge group, and by \cite{Gaiotto:2008ak} for non-Abelian
gauge groups. 

Let us choose generators $S, T$ of $SL(2,\bZ)$:
\beq
S=\left(
\begin{array}{cc}
   0 & 1\\
  -1 & 0 
\end{array}
 \right),
\quad
T=\left(
\begin{array}{cc}
   1 & 1\\
  0 & 1 
\end{array}
 \right).
\label{STdef}
\eeq
An arbitrary element $\varphi$ of $SL(2,\bZ)$ can then be written as
\beq
\varphi=T^{n_1} S T^{n_2} S T^{n_3} \ldots S T^{n_k}.
\label{varphiST}
\eeq
It is therefore sufficient to identify the action of $T$ and $S$. 

It is not difficult to describe the effect of $T$ transformations.
$T^k$
maps $\tau$ to $\tau+k$, which means the shift of the $\theta$-angle by
$2\pi k$. When we have a Janus configuration with $\theta$-angle profile
given by
\beq
\theta(x_3)=\begin{cases}
2\pi k & (x_3>0), \\
0 & (x_3<0),
\end{cases}
\eeq
this induces a level $k$ Chern-Simons term on the 3d domain wall at $x_3=0$:
\beq
 \frac{1}{8\pi^2} \int_{\bR^3 \times \bR}  \!\theta(x_3)\,\textrm{Tr}\,
 F\wedge F=\frac{k}{4\pi}\int_{\bR^3} \textrm{Tr}\, A\wedge \left(dA+\frac{2}{3}A^3 \right),
\eeq
Hence $T^k$ corresponds to adding a level $k$ Chern-Simons term for the
background gauge field. 

The effect of $S$-operation is more subtle. 
For the moment let us set the parameter $m$ to be zero, i.e., the 4d
theory is the $\scN=4$ theory, and discuss the mass deformation later. 

To describe S-operation, let us introduce a 3d theory $T[SU(2)]$. This is a theory on the duality domain
wall for $\varphi=S$. 
This is a 3d $\scN=4$ SQED with two electron hypermultiplets. This theory
has been extensively studied in the context of 3d mirror symmetry
\cite{Intriligator:1996ex}, and is self-mirror.

In $\scN=2$ language, we have a vector multiplet $A$, an adjoint (i.e. neutral)
hypermultiplet $\Phi$, and a set of hypermultiplets 
$\phi_1, \phi_2, \tilde{\phi}_1, \tilde{\phi}_2$ of charge $+1,+1,-1,-1$. 
The theory has R-symmetry
$\textrm{Spin}(4)=SU(2)_N\times SU(2)_R$, and a pair of hypermultiplets
$(\phi_i, \tilde{\phi_i})$ is a doublet of $SU(2)_N$.
The theory has a Lagrangian
\begin{equation}
\begin{split}
\scL=\int\! d^4\theta  \left[\frac{1}{g^2}\left(-\frac{1}{4}\Sigma^2+\Phi^{\dagger} \Phi \right)
+ \left(\phi_i{}^{\dagger} e^{-2V}  \phi_i
+\tilde{\phi}_i{}^{\dagger} e^{2V}  \tilde{\phi}_i
 \right)
\right]
%
+ \int \! d^2\theta \left[ \sqrt{2} \tilde{\phi_i} \Phi \phi_i + \textrm{c.c.}\right],
\end{split}
\end{equation}
where $\Sigma$ is a linear multiplet containing the field strength of
the vector multiplet.
This theory has global symmetry $SU(2)\times SU(2)^{\vee}$,
where $SU(2)^\vee=SO(3)$ is the Langlands dual of $SU(2)$. 
The $SU(2)$ symmetry manifest in the action, and $\phi_i$
($\tilde{\phi_i}$) transform as a fundamental (antifundamental) of that $SU(2)$.
The $SU(2)^\vee$  symmetry is not present in the classical action, and
arises as a quantum symmetry of the theory (the Abelian part of
$SU(2)^{\vee}$ is simply a shift of the dual photon, but we need to use
monopole operators to see other symmetries).

We can introduce a real mass parameter $\mu$ and a FI parameter $\zeta$.
The former is introduced by gauging the $U(1)$ part of the flavor
symmetry $SU(2)$ and introducing a background gauge field 
$V_{\mu}=-i\theta \bar{\theta}\mu$. Since the four hypermultiplets
$\phi_1, \phi_2, \tilde{\phi}_1, \tilde{\phi}_2$ have charges 
$+1, -1, -1, +1$, this adds a term
\beq
\scL_{\mu}=\int \!d^4\theta \left[
\phi_1^{\dagger} e^{-2V_{\mu}} \phi_1 +
\phi_2^{\dagger} e^{2 V_{\mu}} \phi_2 +
\tilde{\phi}_1^{\dagger} e^{2V_{\mu}} \tilde{\phi}_1 +
\tilde{\phi}_2^{\dagger} e^{-2V_{\mu}} \tilde{\phi}_2
\right].
\label{FI}
\eeq
to the Lagrangian\footnote{This parameter $\mu$ is the real mass
parameter, which is one of the triplet of mass parameters. Only the real mass
parameter is consistent with the symmetries used for the localization of
the partition function. The same remark applies to the FI parameters.}.
The FI parameter is similarly introduced by coupling a background vector multiplet
\beq
\scL_{\zeta}=-\frac{4}{\pi} \int \! d^4 \theta \,V_{\zeta}\Sigma,
\eeq
where $V_{\zeta}=i \theta \bar{\theta} \zeta$ and the numerical factor
in front of the integral
is chosen for later convenience.

\bigskip

Now let us go back to the action of $S$ on 3d $\scN=4$ gauge theories.
Let us begin with a 3d superconformal theory with a $SU(2)$ global
symmetry\footnote{More precisely, this means that we have a precise
recipe to couple the theory with a background vector multiplet.}. Since $T[SU(2)]$ has global symmetry 
$SU(2)\times SU(2)^{\vee}$, we can gauge the diagonal part of the two
$SU(2)$ global symmetries of the two theories. The resulting theory has
a global symmetry $SU(2)^{\vee}$, which is a remnant of the global
symmetry of $T[SU(2)]$. Similarly, when we have a theory with global
symmetry $SU(2)^{\vee}$, we can couple it to $T[SU(2)^{\vee}]$ (which is
$T[SU(2)]$ due to mirror symmetry) and the
resulting theory has a global symmetry $SU(2)$.
This is the action of $S$ on 3d theories.

$S$ and $T$ defined above satisfy 
\cite{Witten:2003ya,Gaiotto:2008ak}\footnote{There are some subtleties
in these formulas --- in fact, $S^2$ changes the orientation of the
current and therefore it is more legitimate to write $S^2=-1$. Also,
$(ST)^3$ is in general a certain topological invariant which is
independent of the conformal field theory under study. We are not going
to deal with such subtleties in this paper. Let us remind the readers,
however, that this is simply an overall constant whereas our partition
function is a function of puncture parameters $\{m_i\}$.}
\beq
S^2=1,\quad (ST)^3=1,
\eeq
and hence generate the full $SL(2,\bZ)$.

\begin{figure}[htbp]
\centering{\includegraphics[scale=0.45]{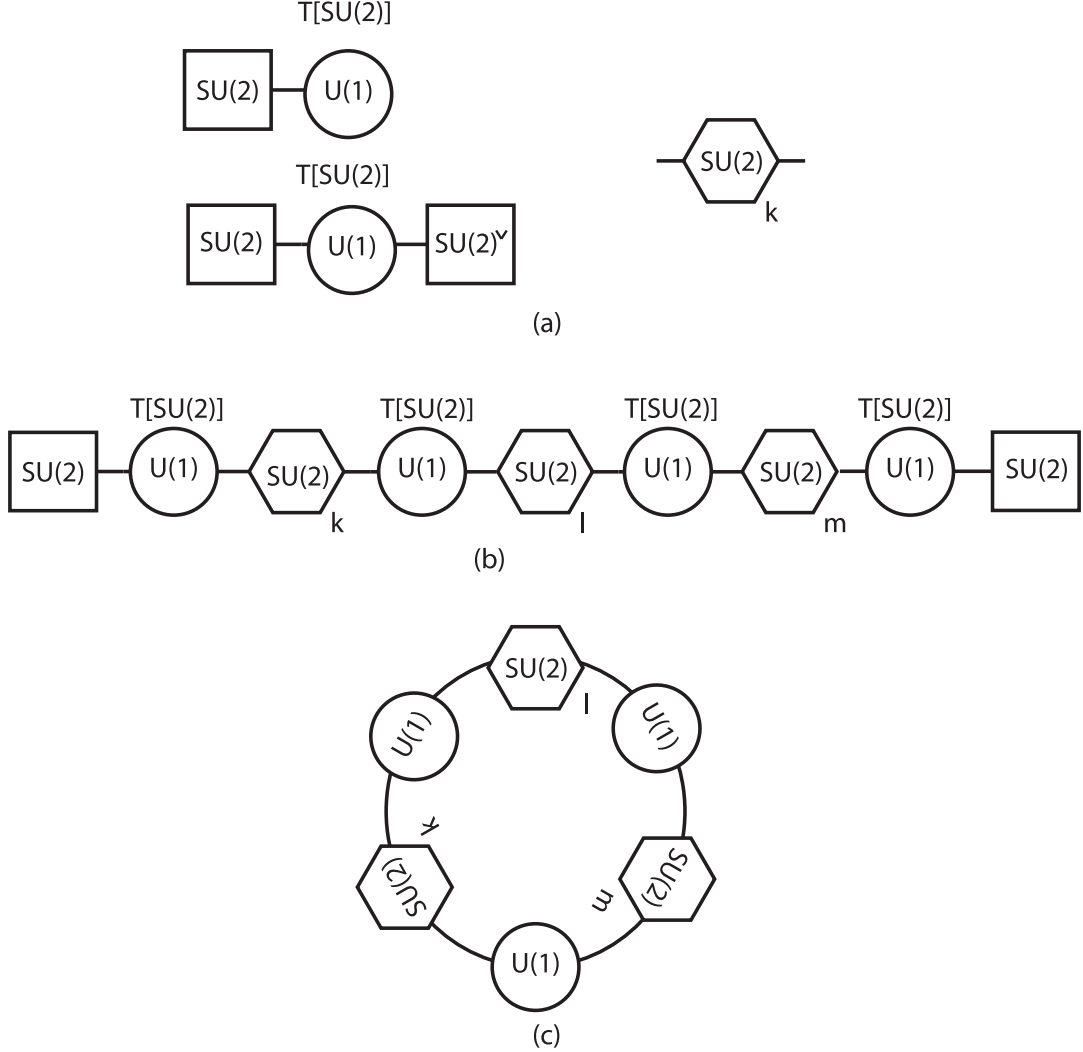}}
\caption{Graphical representation of $S$ and $T$ operations on 3d
 $\scN=4$ theory. In these figures a square represents a global
 symmetry, a circle gauge symmetry, and a hexagon a Chern-Simons term, with level
 specified by an integer written nearby. In (a) we listed all the basic
 building block of our quiver. The left represents our $T[SU(2)]$
 theory. This is often written as above, with only $SU(2)$ symmetry
 manifest. However, it also has quantum $SU(2)^{\vee}$ symmetry, and in
 the figure below we have also shown this symmetry explicitly.
The
 figure on the right represents the action of T, i.e., Chern-Simons
 couplings.
 More generally, an element
 of the mapping class group is represented by a product of $S$ and $T$,
 and graphically represented by a linear quiver. The example of
 $\varphi=ST^k ST^l ST^m S$ is given in (b). When we gauge the remaining
 $SU(2)\times SU(2)$ symmetry, we have a circular quiver, given in (c).
In figures (b) and (c) we neglected subtle differences between $SU(2)$
 and $SU(2)^\vee$.}
\label{STfigure}
\end{figure}

It is useful to graphically represent $S$ and $T$ as in
Figure \ref{STfigure}. In this figure, gauging the diagonal $G$
symmetry ($G=SU(2)$ or $SU(2)^\vee$) of two theories with $G$ global
symmetry is represented by concatenating two edges.
Our 3d theory corresponding to an element of the mapping class groups is described by a linear quiver,
while
the 3d theory in \eqref{main2} is described by a circular
quiver.

\bigskip

Let us now give mass to the adjoint hypermultiplet, and deform $\scN=4$
theory to $\scN=2^*$ theory. The corresponding deformation in 3d daughter theory
was identified in \cite{Hosomichi:2010vh} based on symmetry arguments:
the deformation should preserve $SU(2)\times SU(2)^{\vee}$ symmetry as
well as $\scN=2$ supersymmetry.
The answer is that we weakly gauge the $U(1)$ symmetry under which 
$\phi_i, \tilde{\phi}_i$ all have charge $1$ and $\Phi$ charge $-2$.
The action now contains an extra term
\beq
\scL_{m}=\int \! d^4\theta \left[ \Phi^{\dagger} e^{4V_{m}}\Phi+ \sum_{i=1}^2
\left( \phi_i^{\dagger} e^{-2V_m} \phi_i +\tilde{\phi}_i^{\dagger} e^{-2
V_m} \tilde{\phi}_i \right) \right],
\eeq
where $V_{m}=i m \theta \bar{\theta}/2$. The factor 2 is chosen
for notational simplicity of the following discussion.

\bigskip
\noindent 
{\it Partition Function on Deformed $S^3$}

We are now going to compute the partition function on $S^3$, or more
generally the deformed 3-sphere $S^3_b$.
We can use localization \cite{Kapustin:2009kz}, as well as its
 generalization to theories with anomalous dimensions
\cite{Jafferis:2010un,Hama:2010av}, to compute the partition function on
$S^3$. More recently, these computations has been generalized to the
deformed $S^3_b$ \cite{Hama:2011ea}. 

The partition function for the mass-deformed $T[SU(2)]$, given in
\cite{Hosomichi:2010vh,Hama:2010av} for $b=1$ and here generalized to
$b\ne 1$, reads\footnote{Here we shifted by mass by $iQ/2$ as explained in \cite{Okuda:2010ke}.}
\beq
Z_{T[SU(2);m]}(\mu,\zeta;m)=\frac{1}{s_b(m)}\int\! d\sigma \,\frac{s_{b}(\mu+\sigma+\frac{m}{2}+\frac{iQ}{4}) s_{b}(\mu-\sigma+\frac{m}{2}+\frac{iQ}{4})}  {
s_{b}(\mu+\sigma-\frac{m}{2}-\frac{iQ}{4}) s_{b}(\mu-\sigma-\frac{m}{2}-\frac{iQ}{4}) }\, e^{4\pi i \zeta \sigma}.
\label{Skernelgauge}
\eeq
In this expression the four dilogarithms arise from the 1-loop
determinants of four hypermultiplets.
 The integration variable $\sigma$ is a scalar of the vector
multiplet, and the exponential term in the
integrand represents the classical contribution from the vector
multiplet.
1-loop determinant is trivial for the vector multiplet. 
Finally, the factor $1/s_b(m)$ arises from the 1-loop determinant for 
the neutral hypermultiplet.
Mirror symmetry can be formulated as a statement
\beq
Z_{T[SU(2);m]}(\mu,\zeta;m)=
Z_{T[SU(2);m]}(\zeta,\mu;-m).
\eeq

We can also generalize this analysis to a linear quiver, i.e. more
general element of $SL(2,\bZ)$ as in Figure \ref{STfigure} (b). 
When we gauge $SU(2)$ global symmetry ($SU(2)^\vee$ symmetry), 
the background vector fields represented by the mass parameter $\mu$ (FI
parameter $\zeta$) becomes dynamical, and should be regarded as a
scalar component $\sigma$ of the vector multiplet. 
In localization, this is the variable over which we integral out. 
This means that when we take a product of two elements in $SL(2,\bZ)$,
we need to take a product of the partition function of the corresponding two
theories and integrate over the scalar component of the vector multiplet
which becomes dynamical. For example, for $\varphi=ST^k S$ we have 
\beq
\int \!d\sigma \, Z_{T[SU(2);m]}(\mu,\sigma)\, e^{-4\pi ik \sigma^2}\, Z_{T[SU(2);m]}(\sigma, \mu'),
\label{Zeg}
\eeq
where the term $e^{-4\pi i k \sigma^2}$ represents the classical contribution from the
Chern-Simons term induced by the action of $T^k$.
When we introduce integral kernels
\beq
S_{(\sigma,\sigma';m)}:=Z_{T[SU(2);m]}(\sigma,\sigma';m), \quad
T_{(\sigma,\sigma')}:=e^{-4\pi i \sigma^2} \delta(\sigma-\sigma'),
\label{Tkernelgauge}
\eeq
the expression \eqref{Zeg} then simplifies to 
\beq
\int \! d\sigma d\sigma' \, S_{(\mu,\sigma;m)} T^k_{(\sigma,\sigma')} S_{(\sigma',\mu';m)}=(ST^kS)_{(\mu,\mu';m)}.
\eeq
Namely, $\varphi$ is simply a product of matrices
given in \eqref{Tkernelgauge}! It is therefore straightforward to
compute the partition function for the theory $T[SU(2);\varphi;m]$ for
any element of $\varphi$ of $SL(2,\bZ)$.

This statement, despite its simplicity, is actually far from trivial.
The potential source of trouble is that (as mentioned already) the classical Lagrangian of
$T[SU(2);m]$ has only $SU(2)$ symmetry, and $SU(2)^{\vee}$ is a quantum
symmetry which is not present in the Lagrangian. We can switch to the
mirror description to make $SU(2)^\vee$ symmetry manifest, but then
$SU(2)$ symmetry is turned into a quantum symmetry. What this means is
that at the Lagrangian level it is not clear how to gauge both $SU(2)$
symmetry and $SU(2)^{\vee}$ symmetry simultaneously, but such a
simultaneous gauging
of $SU(2)$ and $SU(2)^{\vee}$ is needed for 
constructing theories with longer quivers
including three or more $S$, say
$\varphi=S T^k S T^l S$ \footnote{We thank F.~Benini and D.~Gaiotto for
discussion on this point.}. 
In this paper, motivated by the correspondence with Liouville theory and
Chern-Simons theory, we conjecture that the partition function in these
cases can still be computed from the products of \eqref{Tkernelgauge}.
Alternatively, we can regard our Gauge/Liouville/Teichm\"uller/Chern-Simons
correspondence as a supporting evidence for this conjecture.
It would be interesting to explicitly verify this conjecture.

\subsection{Comparison with Liouville Theory}

Let us next compare out results in 3d gauge theory with the modular
transformation properties of conformal blocks in Liouville theory \eqref{Ftransform}.
Again, an element $\varphi$ of $SL(2,\bZ)$ can be decomposed into a
product of $S$'s and $T$'s, and all we need to do is to write down the
integral kernel for $S$ and $T$. 
For the $T$ transformation, we have
\beq
T_{(\alpha, \alpha';E)}=\delta(\alpha-\alpha') e^{4\pi i \Delta(\alpha)}, 
\label{TkernelL}
\eeq
where $\Delta(\alpha)$ is the conformal dimension given in \eqref{Deltaalpha}.
For the S-kernel the expression in \cite{Hosomichi:2001xc,Teschner:2003at} reads
\beq
S_{(\alpha,\alpha',E)}=\frac{2^{3/2}}{s_b(E)}
\int_{\bR} dr \frac{s_b(\alpha'+r+\frac{E}{2}+\frac{iQ}{4}) s_b(\alpha'-r+\frac{E}{2}+\frac{iQ}{4})}
{s_b(\alpha'+r-\frac{E}{2}-\frac{iQ}{4}) s_b(\alpha'-r-\frac{E}{2}-\frac{iQ}{4})} e^{4\pi \alpha r},
\label{SkernelL}
\eeq
where the function $s_b(x)$ is the quantum dilogarithm function \cite{FaddeevVolkovAbelian,FaddeevKashaevQuantum,Faddeev95} defined in the appendix \ref{sec.dilog}.

By comparing the formulas \eqref{Skernelgauge}, \eqref{Tkernelgauge}
and \eqref{TkernelL}, \eqref{SkernelL}, we
see that the two result much under the parameter identification
\beq
r=\sigma,\quad \alpha=\zeta, \quad \alpha'=\mu, \quad E=m.
\eeq
This explicitly verifies the relation \eqref{Z=varphi}.

\subsection{Construction of the Hilbert Space}\label{subsec.construction}

Having established the connection between 3d $\scN=2$ theory and
Liouville theory, we move on to quantum \Teichmuller theory.
Readers unfamiliar with \Teichmuller theory are encouraged to consult appendix \ref{sec.Teichmuller}.

\begin{figure}[htbp]
\centering{\includegraphics[scale=0.3]{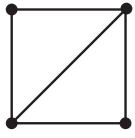}}
\caption{Triangulation of $\Sigma_{1,1}$. The fundamental region of the
 torus is represented by a square, and the puncture are represented by
 four black dots, all of which are identified. The triangulation has two
 triangles and three edges.}
\label{torustriangulation}
\end{figure}

Let us triangulate our surface $\Sigma_{1,1}$ by 
3 edges and 2 triangles, as shown in Figure \eqref{torustriangulation}.
For each of the 3 edges we assign a Fock coordinate, denoted by $x,y$ and
$z$. 
The non-trivial commutation relations are given by \eqref{FockOmega}
\beq
[\sfx,\sfy]=[\sfy,\sfz]=[\sfz,\sfx]=-4 \pi i b^2,
\eeq
or equivalently
\beq
\sfX \sfY=q^{-4}\sfY \sfX, \quad \sfY \sfZ=q^{-4}\sfZ \sfY, \quad
\sfZ\sfX=q^{-4}\sfX \sfZ,
\label{XYZalg}
\eeq
where in the following capitalized variables represent exponentiation, 
\beq
\sfX=e^{\sfx}, \quad \sfY=e^{\sfy}, \quad \sfZ=e^{\sfz}.
\eeq
This algebra has a central element, i.e. a constant, corresponding to the puncture
\beq
m:=\sfx+\sfy+\sfz.
\eeq
There are 2 remaining variables, which is consistent with the fact that
the complex dimension of the \Teichmuller space is one in this example
(see \eqref{dimTgh}).
Let us choose the 2 variables to be $\mathsf{r}$ and $\mathsf{s}$,
which are defined by
\beq
2\pi b \,\mathsf{r}=\frac{\sfy+\sfz}{2}, \quad 
2\pi b \,\mathsf{s}=\frac{\sfy-\sfz}{2}.
\eeq
Note that this particular choice of variables breaks the
$SL(2,\bZ)$ symmetry of the torus.
They satisfy the standard commutation relation
\beq
[\sfs,\sfr]=\frac{1}{2\pi i}.
\label{pq}
\eeq
This commutation relation has a standard representation in terms of
coherent state of $q$:
\beq
\sfr|r\rangle=r |r\rangle , \quad \sfs |r\rangle =\frac{1}{2\pi i}\frac{\partial}{\partial r} |r \rangle.
\eeq
This is an infinite dimensional representation, reflecting the fact the Liouville theory is an irrational conformal field theory.

\bigskip
Let us next describe the action of the mapping class group $SL(2,\bZ)$.
For this purpose it is useful to choose the generators
\beq
L=\left( \begin{array}{cc}
  1 & 1\\
  0 & 1
\end{array}
\right) ,\quad
R=\left( \begin{array}{cc}
  1 & 0\\
  1 & 1
\end{array}
\right).
\eeq
More concretely, these flips change the $\alpha,\beta$-cycles of the torus as
\beq
L: \alpha \to \alpha+\beta, \beta\to \beta, 
\quad
R: \alpha \to \alpha, \beta\to  \alpha+\beta.
\eeq
These generators are related to the generators $S$ and $T$ \eqref{STdef} by
\beq
S=L R^{-1} L ,\quad
T=L.
\eeq

\begin{figure}[htbp]
\centering{\includegraphics[scale=0.4]{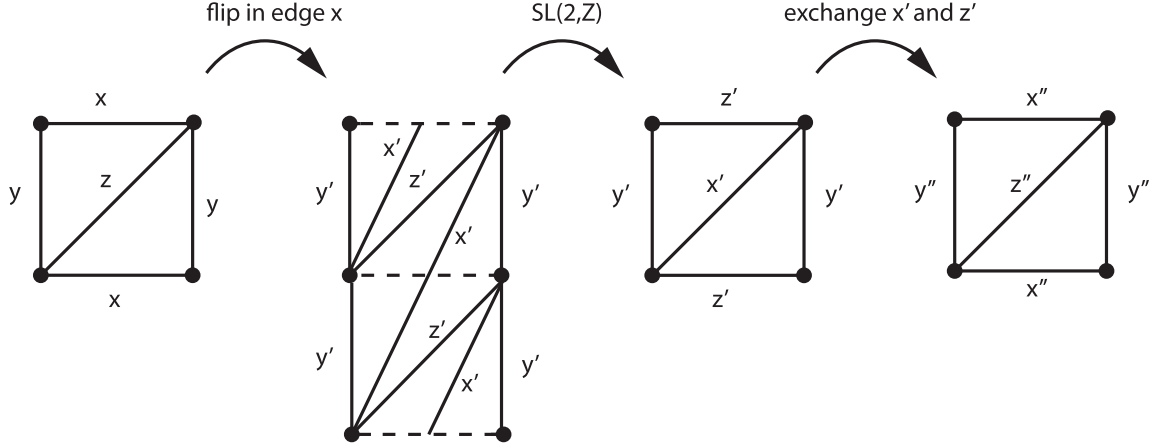}}
\caption{$L$ acts on the triangulation by a combination of a flip with
 an exchange of labels of edges.}
\label{flipLR}
\end{figure}

The action of $L$ can be considered as a product of the flip on the edge
$x$, together with the exchange of labels of $x$ and $z$ afterwards, see
Figure \eqref{flipLR}.
The flip is represented by \eqref{FockQFlip}
\beq
\sfX'=\sfX^{-1}, \quad
\sfY'=(1+q\sfX)(1+q^3 \sfX)\sfY, \quad
\sfZ'=(1+q\sfX^{-1})^{-1} (1+q^3 \sfX^{-1})^{-1} \sfZ.
\eeq
The variables $\sfX', \sfY', \sfZ'$ satisfy
\beq
\sfX' \sfY'=q^{4} \sfY' \sfX', \quad 
\sfY' \sfZ'=q^{4} \sfZ' \sfY', \quad
\sfZ' \sfX'=q^{4} \sfX' \sfZ'. \quad
\eeq 
We need to exchange $\sfX$ and $\sfZ$ in order to go back to the
original algebra \eqref{XYZalg}. 
We then have an expression for the action of $L$:
\beq
\sfX''=(1+q\sfX^{-1})^{-1} (1+q^3 \sfX^{-1})^{-1}\sfZ,  \quad
\sfY''=(1+q\sfX)(1+q^3 \sfX)\sfY, \quad
\sfZ''=\sfX^{-1}.
\label{Laction}
\eeq
The operator $\sfL$ for an element $L$ of the mapping class group
reproduces this
\beq
\mathsf{X}''=\mathsf{L}^{-1}\, \mathsf{X}\, \mathsf{L} ~~, \quad
\mathsf{Y}''=\mathsf{L}^{-1}\, \mathsf{Y}\, \mathsf{L} ~~, \quad
\mathsf{Z}''=\mathsf{L}^{-1}\, \mathsf{Z}\, \mathsf{L} ~~, \quad
\eeq
and we have a similar set of equations for $\sfR$.

\subsection{Chern-Simons Reformulation} \label{sec.CS}

Armed with the construction of the Hilbert space in the previous section,
we can now give a precise prescription to define the Chern-Simons
partition functions in \eqref{main1}, \eqref{main2}.
Here we concentrate on the trace $\textrm{Tr}\,\varphi$, 
but the same argument can straightforwardly be adopted for
 $\langle l | \varphi | l'\rangle$.

Again, let us fix a triangulation of $\Sigma$. The element $\varphi$ of the
mapping class group maps this triangulation to another. Since
any two triangulations are related by flips, the action of 
$\varphi$ on the Hilbert space can be represented by a product of 
operators representing flips.
\beq
\varphi=\sfT_1 \sfT_2 \ldots \sfT_N,
\label{varphidecomp}
\eeq
where in the example here, $\sfT_i$ is either one of the
two operators $\sfL$ or $\sfR$.
The decomposition of $\varphi$ into flips is not unique, but the
expression 
\eqref{varphidecomp} is well-defined due to the pentagon relation.
By inserting a complete set of basis
\beq
\int \!dq_i \,|q_i \rangle \langle q_i |=1.
\eeq
we have
\beq
\textrm{Tr}(\varphi)=\int \prod_i dq_i \, \langle q_1 | \sfT_1 | q_2  \rangle
 \langle q_2 | \sfT_2 |q_3  \rangle \ldots  \langle q_N | \sfT_N |q_1 \rangle.
\eeq
As for the basis $|q_i\rangle$, we can either take $|x\rangle$ or its
conjugate $|y\rangle$.
Note that the expression above is
independent of the choice of basis $|q_i\rangle$, 
since we are computing the trace.

This discussion so far can be summarized in the following rules:

\begin{enumerate}

\item
Decompose the action of $\varphi$ into a product of flips.

\item For each flip we prepare a ``wave function'' $\langle q|\sfT|q'
\rangle$. Here $q$ and $q'$ are different only in quadrilateral where
the flip $\sfT$ takes place. In other words, this is a local operation on 
$\Sigma$.

\item
We glue the wave functions, where gluing means we integrate over the
variables shared by two flips. For example, the gluing of
two wavefunction $\langle q | \sfT| q'\rangle$ and $\langle q'| \sfT'|
q''\rangle$
is performed as
\beq
\int\! dq' \, \langle q | \sfT| q'\rangle \langle q'| \sfT'|q''\rangle,
\eeq
and this gives another wavefunction $\langle q |\sfT \sfT' |q''\rangle$.
By repeating this procedure, the trace $\textrm{Tr}(\varphi)$ is obtained by integrating over
all the intermediate states $|q\rangle$.
\end{enumerate}

Note that the trace is independent of the choice of triangulation we
started with. Indeed, when we change the triangulation we change the
initial state $|q_1\rangle$ to $|q_1'\rangle=\sfT |q_1\rangle$, where
$\sfT$ represents a product of flips. This has the effect of replacing
$\varphi$ by $\sfT^{-1} \varphi \sfT$, but this does not change the
trace since the trace is invariant under conjugation.

\begin{figure}[htbp]
\centering{\includegraphics[scale=0.20]{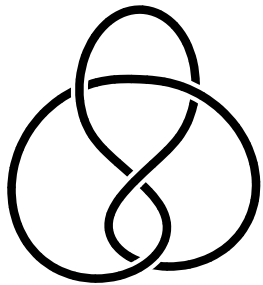}}
\caption{The figure eight knot. This is a knot inside $S^3$, and 
the complement of this knot is a mapping torus of $\Sigma_{1,1}$ over
 $S^1$, with $\varphi=LR$.}
\label{knotfigure}
\end{figure}

\bigskip
Now let us reformulate these rules in terms of 
the 3d $SL(2,\bR)$ Chern-Simons theory. 
Recall that our 3-manifold relevant for the computation of
$\textrm{Tr}\,\varphi$ is a mapping
torus \eqref{mappingtorus}.
In most of the cases the 3-manifold is a hyperbolic manifold, 
locally isomorphic to $\bH^3$. 
For example, for $\varphi=LR$ we have a complement of
the figure eight knot (also called $4_1$ knot, see Figure
\ref{knotfigure}), $\varphi=L^2R$ is
a hyperbolic manifold called m009 in the conventions of SnapPea. 

\begin{figure}[htbp]
\centering{\includegraphics[scale=0.3]{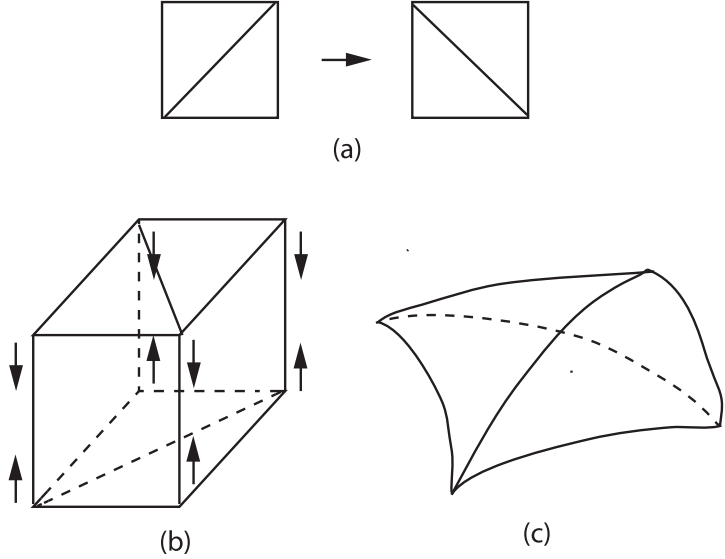}}
\caption{A flip in 2d can be traded for a tetrahedron in 3d. The
 tetrahedron (c) is obtained by a deformation retract of the cube (b)
 representing the flip (a).}
\label{pillow}
\end{figure}

The key observation for the following discussion is that a flip can be traded for a tetrahedron
\cite{FloydHatcher}, see Figure \ref{pillow}. Suppose that we have a flip in a quadrilateral. We
can then introduce an extra direction (time coordinate) to represent
this change as a cube, see Figure \ref{pillow} (b). By taking a deformation
retract of this cube, we have a tetrahedron of the pillow-like shape
(Figure \ref{pillow} (c)).
This tetrahedron is essentially the
superposition of the two quadrilaterals in Figure \ref{pillow} (a).

This means that the decomposition of $\varphi$ into flips in 2d now becomes a
decomposition of the 3-manifold into tetrahedra in 3d language. 
The gluing operation in the third step of the previous rule then 
becomes the gluing two faces of the tetrahedra. 

In our example of once-punctured torus, when $\varphi$ is written as
product of $N$ generators $L$ and $R$, the mapping torus 
is triangulated with 
$N$ tetrahedra. We can take these tetrahedra to be ideal tetrahedra,
where an ideal tetrahedron is a tetrahedron with vertices at the boundaries of
$\bH^3$ in a hyperbolic manifold \cite{FloydHatcher}\footnote{This
particular ideal triangulation coming from 2d triangulation is called the canonical ideal triangulation
in the literature.}.

Our new rule, now in 3d language, is summarized as follows.

\begin{enumerate}
\item
Decompose the mapping torus into tetrahedra
\beq
M=\bigcup_i \Delta_i,
\label{Mdecomp2}
\eeq 

\item
For each tetrahedron we prepare a matrix element 
$\langle q|\sfT|q'\rangle$. 
$|q\rangle$ and $|q'\rangle$ now represent the boundary conditions at
     the faces of the tetrahedron\footnote{For notational simplicity we used
     the same symbols $q, q'$ for slightly different boundary conditions
     in 3d and in 2d; in 3d they determine the
     boundary conditions at the boundary of a tetrahedron,
whereas in 2d they determine the boundary conditions in
     the whole Riemann surface, not only in the quadrilateral.}.

\item
When we glue two ideal tetrahedra, we glue the two corresponding wave
     functions by integrating over boundary conditions. The partition
     function on $M$ is obtained by gluing all the ideal tetrahedra as in
     \eqref{Mdecomp2}.

\end{enumerate}

\begin{figure}[htbp]
\centering{\includegraphics[scale=0.25]{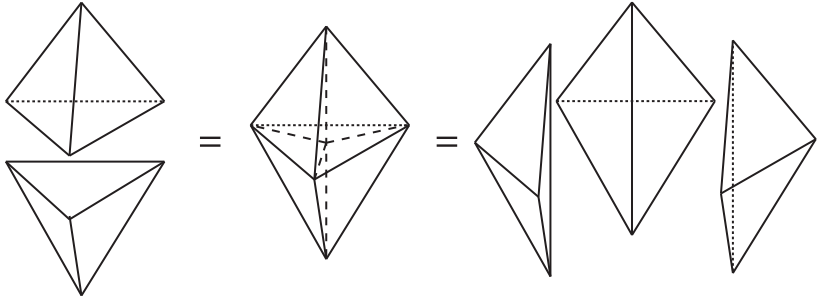}}
\caption{The 2-3 Pachner move. This move changes a 3d ideal triangulation into
 another.}
\label{Pachner}
\end{figure}

Some readers might think at this moment that this is just a trivial
rewriting of what we already know. However, all of the above 3 steps
are given intrinsically in 3d, and in particular, we can choose
triangulations different from the ones coming from the triangulation of
the 2d surface $\Sigma$. Any two 3d ideal triangulations are related by a
series of 2-3 Pachner moves shown in Figure \ref{Pachner}, 
and pentagon
relation in 2d can now be interpreted as an invariance of our partition
function under the 2-3 Pachner move\footnote{This result is known in
the literature, see for example \cite{HikamiGeneralized}. This reference
also discuss examples of once-punctured torus bundles, and computed
their partition functions by using canonical triangulations.}.

The partition function defined by the procedure above is a kind of a state sum model
for the Chern-Simons theory. The basic idea is the same as our previous
discussion of Chern-Simons theory as TQFT, except that we now apply the
same procedure to each tetrahedra. Namely, we have a wave function (an
element in the Hilbert space) for
each tetrahedron when we do the path
integral over the tetrahedron with specific boundary conditions.
There are some well-known state sum models for Chern-Simons theory, see
\cite{TuraevViro,Ooguri:1991ni}. 
For the case of direct relevance to our paper, i.e. $SL(2,\bR)$
Chern-Simons theory or its analytic continuation into $SL(2,\bC)$, there
is a proposed state sum model by Hikami
\cite{HikamiHyperbolic1,HikamiGeneralized}, which is a non-compact
analog of \cite{Kashaev6jsymbol}. 
See also
\cite{Dimofte:2009yn,Dijkgraaf:2010ur,Dimofte:2011gm}.
It is an interesting problem to compare the our state sum model
with
the known results in the literature. This problem is currently under
investigation \cite{InProgress}\footnote{In \cite{Terashima:2011xe}, for
a once-punctured torus bundle, it
was shown that the semiclassical classical limit of the trace of the
\Teichmuller theory reproduces the hyperbolic volume and the
Chern-Simons invariant of the mapping tori.}. 

\bigskip
Finally, let us conclude this section by briefly commenting on more general
Riemann surfaces. The procedure is in principle the same for any Riemann
surface, but can be technically involved. 
Probably the next simplest example will be a 4-punctured sphere, which
corresponds to 4d $SU(2)$ $N_f=4$ theory. 
The integral kernel of the fusion and branding for this theory are
known (see for example \cite{Ponsot:2000mt}). Indeed, the
S-transformation for $\Sigma_{1,1}$ is a specialization of that of
$\Sigma_{0,4}$ \cite{Hadasz:2009sw}. Both $\Sigma_{1,1}$ and
$\Sigma_{0,4}$ are covered by $\Sigma_{1,4}$.

\section{Comments on Higher Rank Generalizations}\label{sec.higher}

It is straightforward to generalize our proposal to the
case $G=SU(N)$. This claims the equivalence of the partition function of
3d $SL(N,\bR)$ Chern-Simons theory on a 3-manifold $\Sigma\times I$ with that of a 
3d $\scN=2$ $SU(N)$ gauge
theory on a (deformed) 3-sphere $S^3$ ($S^3_b$),
which is realized as a duality domain wall inside the mother 4d $\scN=2$ $SU(N)$ theory
on $S^4$. 

Many of the ideas presented in section \ref{sec.idea} have their
counterparts in this higher rank generalization.
For example, AGT relation has been 
generalized to $SU(N)$ gauge groups by \cite{Wyllard:2009hg,Mironov:2009by}. Moreover, there exists are higher rank
generalization of \Teichmuller theory called the higher \Teichmuller theory
\cite{FockGoncharovHigher}, which quantizes the moduli space of flat
$SL(N,\bC)$ or $SL(N,\bR)$ connections for $N>2$
\cite{HitchinTeichmuller}.
As for the boundary conditions of Chern-Simons theory, 
there is again a Hamiltonian reduction from $SL(N,\bC)$
Chern-Simons theory to Toda theory on the boundary \cite{Alekseev:1988ce,Bershadsky:1989mf,Forgacs:1989ac}. 
Moreover, the computation of partition functions by localization works
for $SU(N)$ gauge theories both for 3d and 4d theories.
It would be interesting to perform detailed quantitative checks of this
generalization\footnote{It is probably work mentioning here that a class of higher 
spin theories can be reformulated as a $SL(N,\bR)\times SL(N,\bR)$
Chern-Simons theory. There are recent proposals about holographic
duality with $\scW$-minimal models; see for example \cite{Gaberdiel:2010pz} for a recent discussion.}.

\section{Conclusion and Discussion} \label{sec.discussion}

In this paper, we initiated a program to connect 3d $SL(2,\bR)$
Chern-Simons theory and 3d $\scN=2$ gauge theory on duality walls. In particular,
we proposed an equality of the partition functions of the two theories,
given in \eqref{main1} and \eqref{main2}. We also proposed a $SL(2,\bR)$
reformulation of the AGT relation and its generalization (see Table
\ref{SummaryTable}). We provided evidence for this conjecture by linking
the two theories with quantum Liouville theory and quantum \Teichmuller
theory. We also provided explicit computations in the case of the
once-punctured torus.

Clearly there are many open problems, some of which are already
mentioned in the main text. Here we list some more problems. Solutions
to any of these questions are welcome.

\begin{itemize}

\item Identify the matter content of the 3d $\scN=2$ domain wall theory for
      general Riemann surface $\Sigma_{g,h}$. This should be a doable
      problem since the partition function of these theories can be
      computed from quantum \Teichmuller theory.

\item Can we give a direct verification of our proposal 
from dimensional reduction of 6d $(2,0)$ theory? As a possible clue, \cite{Witten:2009at}
      discuss the appearance of the Chern-Simons term from a dimensional reduction of $(2,0)$ theory.

\item Does our 4d partition function give an invariant of a
      3-manifold determined from a Heegaard decomposition (see section \ref{subsec.reformulation})?
      Alternatively, one may use the definition from the surgery
      formula, mimicking the argument of \cite{Witten:1988hf}.

\item Identify the quantum group $U_q(SL(2,\bR))$ in our story, see for
      example \cite{Ponsot:2000mt}. For
      $SU(2)$ Chern-Simons theory the quantum group $U_q(SU(2))$ was used in the mathematical formulation of
      the 3-manifold invariants \cite{ReshetikhinTuraev}. 

\item $SL(2,\bC)$/$SL(2,\bR)\times SL(2,\bR)$ Chern-Simons theory describe the Lagrangian of 3d gravity
      \cite{Witten:1988hc}. 
      Is there any implication of our results to 3d gravity?
    Since Liouville theory describes 2d gravity, there should be a relation
      between 3d gravity and 2d gravity, which in turn is related to
      4d/3d supersymmetry gauge theory.

\item Extend our results to asymptotically free 4d gauge theories \cite{Gaiotto:2009ma,Marshakov:2009gn} or
      5d gauge theories \cite{Awata:2009ur,Awata:2010yy}.

\item Study the self-duality of $SL(2,\bR)$ Chern-Simons theory, see
      discussion around \eqref{bdual} and \cite{Dimofte:2011gm} for
      recent discussion. Mathematically, this is a manifestation of the
      (quantum ) geometric Langlands correspondence, see for example
      \cite{Stoyanovsky:2006mj,Tan:2007ej,Tan:2008ak,FrenkelLectures,Giribet:2008ix,Teschner:2010je}.

\item 
      Formulas of quantum dilogs appear in the
      Kontsevich-Soibelman wall crossing formula
      \cite{KontsevichSoibelamn}. Is there any implication of our
      results to wall crossing? The reference \cite{Gaiotto:2009hg}
      uses Fock coordinates to give physical derivation of the wall
      crossing formula, and 
      the trace similar to ours appear in \cite{Cecotti:2010fi}.

\item Prove \eqref{topologicalphysical}. Connect our story to that
      of \cite{Nekrasov:2009rc,Nekrasov:2010ka,Teschner:2010je,Bonelli:2009zp}, and 
      find integrable structures in Chern-Simons theory.

\item Find a gravity solution representing our M5-brane
      configurations in the large $N$ limit. The gravity solution
      constructed in \cite{Gauntlett:2000ng} may be useful in this respect.

\item Reformulate/prove our conjectures in the language of Penner-type
      matrix models, see \cite{Dijkgraaf:2009pc}.

\item The dual graph of the triangulation of $\Sigma$ is a bipartite
      graph (dimer), which also appears in the context of 4d quiver gauge
      theories \cite{Hanany:2005ve,Franco:2005rj,Franco:2005sm} and BPS state counting on toric Calabi-Yau
      manifolds \cite{Szendroi:2007nu,Mozgovoy:2008fd,Ooguri:2008yb}.
      There are similarities between quantum \Teichmuller theory and dimer theory, cf. \cite{KenyonGoncharov}.
      How far does this analogy go? 
 
\end{itemize}

\section*{Acknowledgments}

This work originated from discussion during the workshop 
``Encounter with Mathematics'', Chuo University, May 2008, 
and we would like to thank the organizers for providing stimulating
environment. M.~Y. would like to thank 
Aspen Center for Physics 
for hospitality, where part of this work has been performed.
M.~Y. would like to thank Akishi Kato for discussion on hyperbolic geometry and knot theory in 2005, 
which has provided him with the underlying motivation for this work.
Special thanks goes to T.~Dimofte for discussion on the Chern-Simons
theory and the contents of section \ref{sec.CS}.
We would also like to thank A.~Goncharov, F.~Benini,
H.~Fuji, D.~Gaiotto, S.~Gukov, T.~Hartman, L.~Hollands, K.~Hosomichi, T.~Nishioka, J.~Song, P.~Su{\l}kowski,
J.~Teschner, H.~Verlinde and E.~Witten
for helpful comments, correspondence and stimulating discussion.
The research of M.~Y.~ is supported in part by Princeton Center for
Theoretical Science, 


\appendix

\section{Construction of the \Teichmuller Hilbert Space} \label{sec.Teichmuller}

In this section we summarize basic aspects of the classical and quantum \Teichmuller
theory. This gives a precise recipe to construct the Hilbert space
$\scH_T(\Sigma)$. Our discussion here applies to a general Riemann
surface $\Sigma_{g,h}$. See section \ref{sec.example} for a concrete
discussion in the case of $\Sigma_{1,1}$.

\subsection{Fock coordinates and Flips}

Let us begin by introducing coordinates in the \Teichmuller space.
The basic idea needed for this is simple --- by triangulating a
Riemann surface, we can divide $\Sigma$ into a set of triangles, and by
explicitly specify how we glue these triangles back we can parametrize the 
complex structure moduli of $\Sigma$.

The triangulation is chosen in such a way that 
all the vertices of the triangles are placed at the punctures and all
edges are geodesics connecting punctures\footnote{The dual of this triangulation is called a fat graph.}. Here geodesic means geodesic in the metric corresponding to a point in the \Teichmuller space. 
The number of faces ($F$), edges ($E$) and vertices ($V$) are given by
\beq
  F=2(2g-2+h) , \quad E= 3(2g-2+h), \quad V= h.
\label{VEF}
\eeq
To verify this, note the constraints
\beq
3F=2E, \quad \chi(\Sigma_{g,0})=-V+E-F=2g-2.
\eeq

\begin{figure}[htbp]
\centering{\includegraphics[scale=0.3]{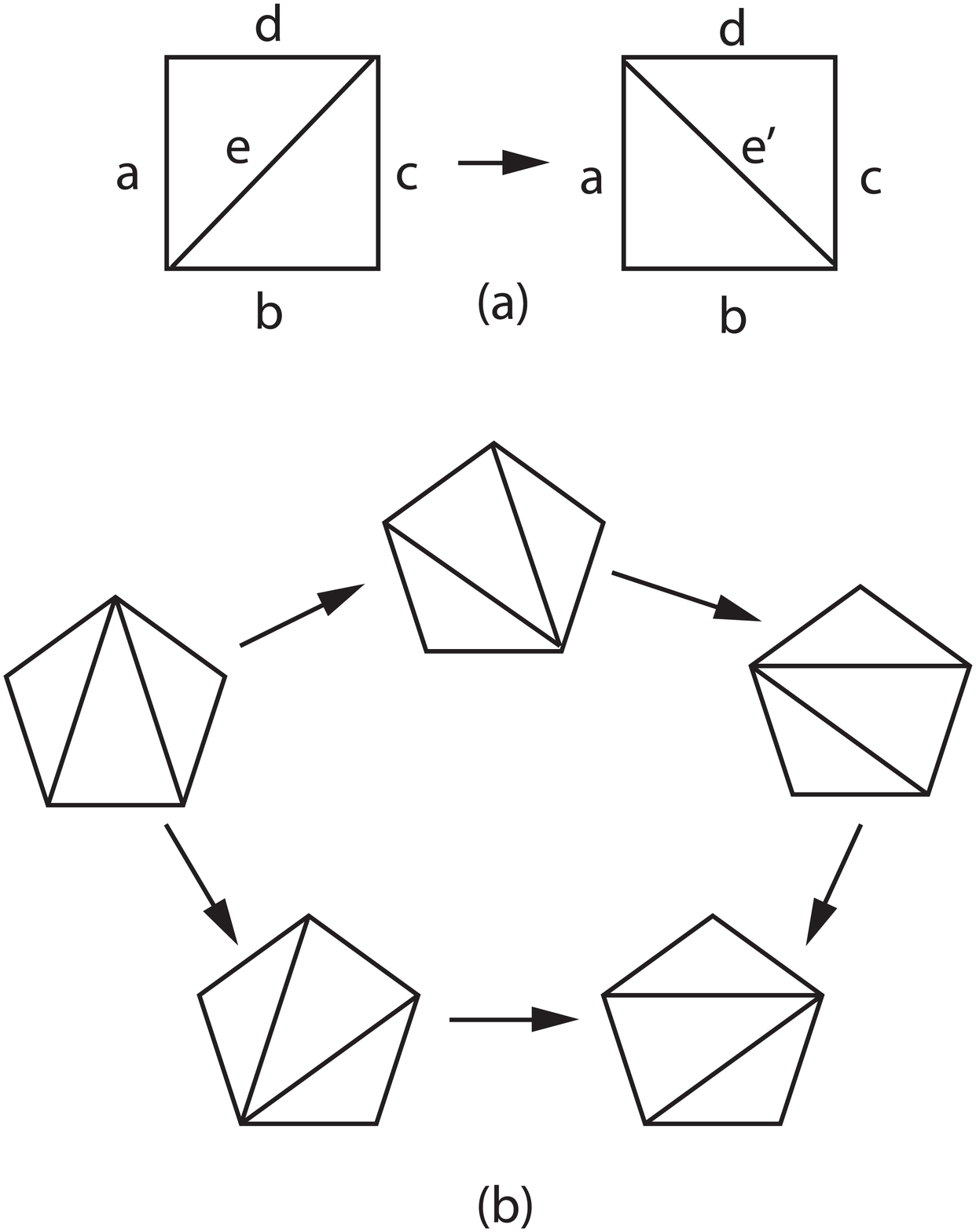}}
\caption{(a) represents a flip on the edge $e$. (b) shows the pentagon relation.}
\label{flip}
\end{figure}

It is known that all such triangulation are related by a series of operations called flips (also called Whitehead moves). 
Choose an edge $e$ which bounds two triangles. A flip removes the edge $e$ and adds another diagonal $e'$ of the resulting quadrilateral
(see Figure \ref{flip}). Flips satisfy the pentagon relation shown in
Figure \ref{flip}, which can be represented by
\beq
T_{12} T_{13} T_{23}=T_{23} T_{12},
\label{PennerPentagon}
\eeq
where $T_{ij}$ represents the flip exchanging the two neighboring faces
$i$ and $j$\footnote{A set of invertible operations $T_{ij}$ satisfying \eqref{PennerPentagon} is called a Ptolemy groupoid.}.

\begin{figure}[htbp]
\centering{\includegraphics[scale=0.35]{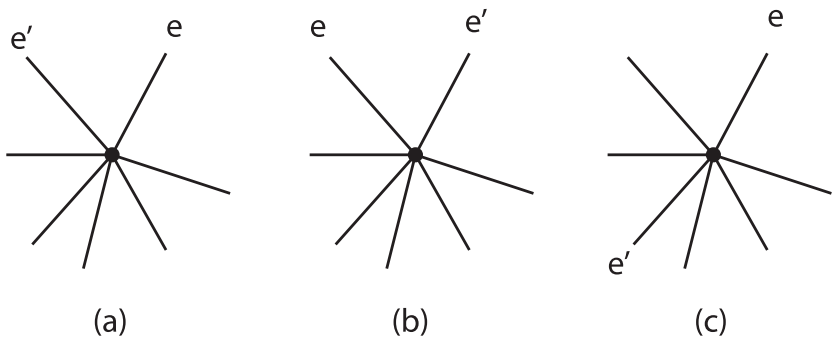}}
\caption{Suppose two edges $e$ and $e'$ share a vertex, and there are no
 other edges between then. This contributes either $1$ or $-1$
  depending on the two possibilities (a) and (b).
 The number $n_{e,e'}$ is defined as a sum of all such
 contributions. Note that (c), with other edges between $e$ and
 $e'$, does not contribute to this computation.}
\label{nee}
\end{figure}

Given a triangulation,  we assign a coordinate $z_e$, the Fock coordinate
(also called the shear coordinate) \cite{Fock}, for each edge $e$. This is a
coordinate of the \Teichmuller space, such that its K\"ahler form
(Weil-Petersson form)
takes a simple form
\beq
\{z_e, z_{e'} \}=n_{e,e'}\in \{-2,-1,0,1,2 \},
\eeq
where the constant $n_{e,e'}$ on the right hand side is defined as the number of times we see Figure \ref{nee} (a) 
minus the number of times we see Figure \ref{nee} (b).

\bigskip

Let us describe the geometrical meaning of the Fock coordinates.
For an edge $e$, let us define $l_e$ to 
be the hyperbolic distance between the two punctured
connected by the edge $e$.
This defines yet another coordinate coordinate of the \Teichmuller
space, called
Penner coordinate or lambda-length \cite{PennerDecorated}.
For an edge $e$ as in Figure \ref{flip} (a), 
the relation between the two coordinates are given by
\footnote{Actually, the definition of $l_e$ is subtle when the puncture has size
zero, since when naively computed $l_e$ diverges. We thus need to choose a
cycle around each puncture and regularize the value of $l_e$.
The Penner coordinate depends on the choice of the regularization, but these
ambiguities cancel out in the Fock coordinates.
}
\beq
Z_e= \frac{L_a\, L_c}{L_b\, L_d},
\label{FockInPenner}
\eeq
where we used capitalized letters when we exponentiate variables. For example,
\beq
L_e=e^{l_e}, \quad Z_a=e^{z_a}.
\eeq
In other words, Fock coordinate is defined as a cross
ratio of Penner coordinates of the four the edges of the
quadrilateral.

\begin{figure}[htbp]
\centering{\includegraphics[scale=0.4]{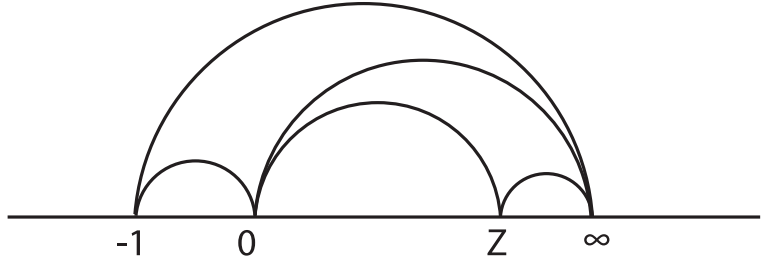}}
\caption{Fock coordinate determines the gluing to two triangles in the
 upper half plane. The vertices on the boundary is taken to be
 $0,-1,\infty$ and $Z$.}
\label{CrossRatio}
\end{figure}

Another way to explain this is as follows.
To explain this, let us
represent the Riemann surface on the upper half plane, and the triangles
as half-circles having their vertices on the boundary. The
uniformization theorem guarantees that by a suitable coordinate
transformation this is always possible for
Riemann surfaces satisfying \eqref{hyperboliccond}. 
By a M\"{o}bius transformation on the boundary we can choose the
vertices of the first triangle to be $0, -1$ and $\infty$. The Fock
coordinate $Z$ then coincides with the coordinate of the fourth vertex
(Figure \ref{CrossRatio}). This means that Fock coordinates parametrizes the gluing to two
triangles.

Let us describe the change of coordinates under the change of
triangulation.
For this, it is sufficient to describe the case of the flip as in Figure
\eqref{flip}. In Penner coordinates, a flip as in Figure \ref{flip} is represented by
\beq
L_e'= (L_a L_c+L_b L_d)/L_e, 
\label{FlipInPenner}
\eeq
while Penner coordinates for all other edges stay the same. This is simply a Ptolemy's theorem in
hyperbolic geometry. 
Using coordinate transformations in
\eqref{FockInPenner},
this is translated into
(assuming the four edges $a, b, c, d$ are all different)
\begin{equation}
\begin{split}
Z_a' = (1+Z_e) Z_a, \quad 
&Z_b'= (1+Z_e^{-1})^{-1} Z_b, \\
Z_c' = (1+Z_e) Z_c, \quad 
&Z_d'= (1+Z_e^{-1})^{-1} Z_d, \\
Z_e'&=Z_e^{-1}.
\label{FockFlip}
\end{split}
\end{equation}
More generally, for an edge $f$ different from $e$ we have
\beq
Z_f'=(1+Z_e)^{n_{f,e}} Z_f \quad (n_{f,e}\ge 0), \quad \quad 
Z_f'=(1+Z_e^{-1})^{n_{f,e}} Z_f \quad (n_{f,e}<0). \quad 
\eeq
We can directly verify that \eqref{FlipInPenner}, \eqref{FockFlip}
satisfy the pentagon relation in Figure \ref{flip} (b).

\bigskip
Finally, let us proceed to the quantum theory. We choose the quantization in
Fock variables. See \cite{KashaevQuantization,KashaevQuantum} for
quantization in other variables defined by Kashaev\footnote{These
variables are assigned to the faces of the triangulation, and may be
useful
for the comparison with the results of
\cite{HikamiHyperbolic1,HikamiGeneralized}}, and see \cite{GuoLiu} the
relation between the two
quantizations.

The quantization proceeds in the standard way: by replacing the
symplectic form by a commutator of operators. For example, in Fock
coordinates we have (recall we have $\hbar=2\pi b^2$ \eqref{kb})
\beq
[\mathsf{z}_e,\mathsf{z}_{e'}]=2\pi i b^2 n_{e,e'},
\label{FockOmega}
\eeq
or equivalently 
\beq
\sfZ_e, \sfZ_{e'}=q^{2 n_{e,e'}} \sfZ_{e'} \sfZ_e,
\eeq
where we defined
\beq
q=e^{\pi i b^2}.
\eeq

The equations for the flip, \eqref{FockFlip}, is now modified to be
\begin{equation}
\begin{split}
\sfZ_a' =(1+q\sfZ_e) \sfZ_a, \quad 
&\sfZ_b'=(1+q\sfZ_e^{-1})^{-1} \sfZ_b, \\
\sfZ_c' =(1+q\sfZ_e) \sfZ_c, \quad 
&\sfZ_d'=(1+q\sfZ_e^{-1})^{-1} \sfZ_d, \\
\sfZ_e'&=\sfZ_e^{-1}.
\label{FockQFlip}
\end{split}
\end{equation}
In the classical limit $q\to 1$ this reduces to the classical
formula for quantization.
We can directly verify that this satisfies the pentagon relation.
Moreover, \eqref{FockQFlip} preserves the relation
\eqref{FockOmega}\footnote{Note that the value of $n_{e,e'}$ in general
changes 
before and after the quantization.}.
In fact, under several assumptions this quantization is unique, as has been shown in \cite{Bai}.

\subsection{Length Operators} \label{subsec.length}

Let us now comment on the construction of the length basis $|l\rangle$ needed for
the construction of conformal block \eqref{FinTeichmuller}\footnote{See
\cite{Teschner:2003at,Drukker:2009id,Teschner:2010je} for the definition of the holomorphic basis
$|\tau\rangle$. Mathematically, this is intimately connected with the
theory of opers \cite{BeilinsonDrinfeldOpers}. The normalizations of the
basis $|\tau\rangle$ is fixed by imposing the asymptotic boundary
conditions for the conformal blocks.}.

For this, let us fix a pants decomposition of $\Sigma$. This means 
$\Sigma$ can be viewed as a union of trinions (a sphere with 3 holes), 
and we connect the trinions by a cylinder.
For each trinion  with the size of holes $l_i$ fixed, it is known that there exists a unique metric with
negative constant curvature. Moreover, when we glue two trinions
together by a cylinder, we can
add a twist $\theta_i$ along the cylinder circle $C_i$ (we call this a pants circle)
along the cylinder. We call this twist parameter. 
The length parameters $l_i>0$ and twist parameters $0\le \theta_i < 2\pi$
(also called Fenchel-Nielsen coordinates)
parametrize the \Teichmuller space of the Riemann surface.
The important properties of these coordinates is that Weil-Petersson form
simplifies in this basis \cite{Wolpert}
\beq
\{l_i, \theta_j\}=\delta_{ij}, \quad \{l_i, l_j\}=\{\theta_i, \theta_j\}=0.
\eeq
This provides an important consistency check of the previous proposal
that $l$ should be identified with the 
label for the Virasoro representation (see \cite{Verlinde:1989ua}).
The shift of the twist angle $\theta_i$ by $\Delta\theta_i$
is represented by an operator $e^{i \Delta\theta_i l_i}$, but this
should also be the same as $e^{i\Delta\theta_i L_0(C_i)}$, where
$L_0(C_i)=\oint_{C_i} dz T(z)$ represents the rotation along the circle
$C_i$.
This immediately means $l(C_i)=L_0(C_i)$.

\begin{figure}[htbp]
\centering{\includegraphics[scale=0.3]{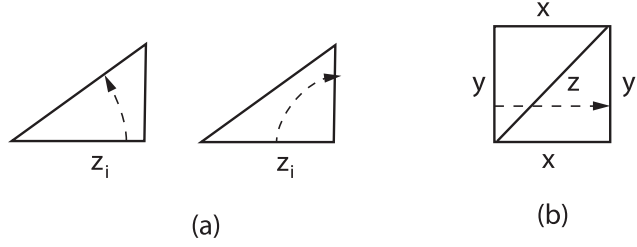}}
\caption{In (a) we list two possibilities for the paths $\gamma_i$. In
 (b) we draw the $\alpha$-cycle on torus, from we read off the classical
 expression of the geodesic length of the $\alpha$-cycle.}
\label{length}
\end{figure}

Let us summarize the classical expression for the geodesic
length in terms of Fock coordinates \cite{Fock}.
Let us choose a geodesic $\gamma$ on $\Sigma$ (in the complex structure
which is kept fixed in this section). We assume that $\gamma$ does not pass
through punctures of $\Sigma$. 
Then $\gamma$ can be described as a series of segments $\gamma_i$, where each
$\gamma_i$ is a path starting from an edge and ending at another edge of
the triangle. Let us denote the Fock coordinate of the edge at the
starting point of $\gamma_i$ by $z_i$.
Depending on the two possibilities shown in Figure \ref{length} (a),
we define 
$M_{\gamma_i}=V E(z_i)$ or  
~~\textrm{or} ~~
$M_{\gamma_i}=V^{-1} E(z_i)$, 
where
the matrices $V, E(z)$ are given by
\beq
V=\left( 
 \begin{array}{cc}
    1 & 1 \\
    -1 & 0 \\
 \end{array}
 \right),
\quad 
  E(z)= 
\left( 
 \begin{array}{cc}
    0  & e^{z/2} \\
    -e^{-z/2} & 0 \\
 \end{array}
 \right), 
\eeq
Then the classical length $l_{\gamma}$ of $\gamma$ is given by the expression
\beq
L_{\gamma}=2\cosh \frac{l_{\gamma}}{2}=\Big|\textrm{Tr} \prod_i M_{\gamma_i}\Big|,
\eeq
where the ordering of the product is determined by the ordering of the
segments $\gamma_i$ inside $\gamma$.

As an example, for once-punctured torus triangulated as in Figure
\ref{torustriangulation}, we can choose an $\alpha$-cycle as in Figure
\ref{length} (b).
\beq
L_{\alpha}=\left|\textrm{Tr} \left(V^{-1} E(z) V E(y)\right)\right|
=2\cosh \left(\frac{y+z}{2} \right)+\exp\left(\frac{y-z}{2}\right).
\label{Lalphaformula}
\eeq

\bigskip

Quantization of the classical geodesic length $L_{\gamma}$ gives a 
length operator $\scL_{\gamma}$ mentioned in the main text\footnote{In general, there are ordering
ambiguities in this quantization, which are fixed such that several
consistency conditions are satisfied. This subtlety does not arise for
the example discussed above.}. The length operators for the pants circles
$\gamma_i$ all commute, since pants circles are non-intersecting. 
This means that we can 
simultaneously diagonalize the operators $\scL_{\gamma_i}$:
\beq
\scL_i | l\rangle =2\cosh (2\pi b l_i)|l\rangle.
\eeq
This is the definition of length basis $| l \rangle$.

For the case of the once-punctured torus,
the length operator for the $\alpha$-cycle $\scL_{\alpha}$ is the quantization of \eqref{Lalphaformula}
\beq
\scL_{\alpha}=2\cosh 2\pi b\sfr+e^{2\pi b \sfs},
\label{Lalpha}
\eeq
see section \ref{subsec.construction} for definitions of $\sfr, \sfs$
and the their coherent representation $|\sfr\rangle$.
The basis $|l \rangle$ is defined as an eigenstate of the length $L_{\alpha}$.
\beq
\scL_{\alpha} |l \rangle =2 \cosh 2\pi b l |l\rangle.
\eeq
Sandwiching \eqref{Lalpha} between $\langle r |$ on the right and $|l\rangle$ on the right, we have a difference equation
\beq
\langle r-ib |l\rangle =(2\cosh 2\pi b l-2\cosh 2\pi r l) \langle r | l \rangle.
\eeq
From which we derive (up to possible overall normalization factors)
\beq
\langle r |l \rangle =\frac{s_b(l+r+iQ/2)}{s_b(l-r-iQ/2)},
\label{wavefunction}
\eeq
where we used the property of the quantum dilogarithm  \eqref{sbdiffeqn}.
We can directly verify the completeness relation \eqref{nucomplete}
by using an identity of quantum dilogarithms, see \cite{KashaevQuantum}.

We can also define $|\tilde{l'}\rangle$, an eigenstate of the
$\beta$-cycle length $\scL_{\beta}$.
The integral kernel for S-transformation \eqref{SkernelL}is then given
by
a pairing between the two
\beq
S_{(l,l')}=\langle l| S |l'\rangle =\langle l |\tilde{l}'\rangle.
\eeq


\section{Quantum Dilogarithm} \label{sec.dilog}

In this appendix we collect formulas for the non-compact quantum
dilogarithm function $s_b(z)$ and $e_b(z)$, which was discovered by
Faddeev and his collaborators
\cite{FaddeevVolkovAbelian,FaddeevKashaevQuantum,Faddeev95}. See also
\cite{VolkovNoncommutative}, section III of \cite{Ruijsenaars} and an appendix of \cite{Kharchev:2001rs}. 

The function $s_b(z)$ is defined by
\beq
s_b(z)=\exp\left[ \frac{1}{i} \int_0^{\infty} \frac{dw}{w} \left( \frac{\sin 2zw}{2\sinh (bw) \sinh (w/b)}-\frac{z}{w} \right) \right].
\label{sbdef}
\eeq
In the literature we also find
\beq
e_b(z)=\exp\left(\frac{1}{4} \int_{-\infty+i0}^{\infty+i0}  \frac{dw}{w} \frac{e^{-i 2zw} }{\sinh(wb) \sinh(w/b)}\right), 
\label{ebdef}
\eeq
where the integration contour is chosen above the pole $w=0$. 
In both these expressions we require $|\mathrm{Im}\, z| <|\mathrm{Im}\, c_b|$ for convergence at infinity. There is a simple relation between the two functions 
\beq
e_b(z)=e^{\frac{\pi i z^2}{2}} e^{- \frac{i\pi (2-Q^2)}{24}}s_b(z),
\label{ebsb}
\eeq
and we loosely refer to both functions as quantum dilogarithms.
The relation \eqref{ebsb} can be shown by decomposing the contour in \eqref{ebdef} into the three parts $[-\infty, -\epsilon] \cup \epsilon e^{[i\pi, 0]} \cup [\epsilon, \infty]$, simplifying expressions, and taking the limit $\epsilon \to 0$.
In the classical limit $b\to 0$, we have
\beq
e_b(z)\to \exp \left(\frac{1}{2 \pi i b^2} \textrm{Li}_2(-e^{2\pi b z}) \right),
\eeq
where $\textrm{Li}_2(z)$ denotes Euler classical dilogarithm function,
defined by
\beq
\textrm{Li}_2(z)=-\int_0^z \frac{\log(1-t)}{t}dt. 
\label{Li2}
\eeq

The quantum dilogarithm function $s_b(z)$ has a number of interesting
properties.  
For example, from the definition it follows immediately that
\beq
s_b(z)=s_{1/b}(z).
\label{sbselfdual}
\eeq
and
\beq
s_b(z) s_b(-z)=1.
\label{sbinverse}
\eeq
For us, it is important that $s_b(z)$ satisfy a difference equation
\beq
s_b\left(z-\frac{ib}{2} \right)=2 \cosh \left(\pi b z\right) s_b\left(z+\frac{ib}{2}\right),
\label{sbdiffeqn}
\eeq
Similarly, for $e_b(z)$ we have
\beq
e_b(z-ib^{\pm 1}/2)=(1+e^{2\pi b^{\pm 1} z}) e_b(z+ib^{\pm 1}/2).
\eeq
We can use these equations to analytically continue 
$s_b(z), e_b(z)$ to the whole complex plane. 
$s_b(z)$ is then a meromorphic function with an infinite product expression 
\beq
s_b(z)=\prod_{m,n\in \bZ_{\ge 0}} \frac{mb+nb^{-1}+\frac{Q}{2}-iz}{mb+nb^{-1}+\frac{Q}{2}+iz}.
\eeq
%
%


\bibliography{AGTknot}
\bibliographystyle{JHEP}


\end{document}